\documentclass{svjour2}                    % onecolumn
\smartqed  % flush right qed marks, e.g. at end of proof
\usepackage{latexsym}

\usepackage{amsmath}    % need for subequations
\usepackage{amsfonts}  %note how statements can be commented out
\usepackage{amssymb}
\usepackage{slashed}

\usepackage{graphicx}
\begin{document}

\title{Quantum Mechanics and the Principle of Least Radix Economy %\thanks{Grants or other notes
}


\author{Vladimir Garcia-Morales          %etc.
}


\institute{Institute for Advanced Study - Technische Universit\"{a}t M\"{u}nchen, Lichtenbergstr. 2a, D-85748 Garching, Germany \\
           \and
           Nonequilibrium Chemical Physics - Physics Department - Technische Universit\"{a}t M\"{u}nchen, James-Franck-Str. 1, D-85748 Garching, Germany
Tel.: +49 89 289 13878\\
\email{vmorales@ph.tum.de}
}

\date{Received: date / Accepted: date}

\maketitle

\begin{abstract}
A new variational method, the principle of least radix economy, is formulated. The mathematical and physical relevance of the radix economy, also called digit capacity, is established, showing how physical laws can be derived from this concept in a unified way. The principle reinterprets and generalizes the principle of least action yielding two classes of physical solutions: least action paths and quantum wavefunctions. A new physical foundation of the Hilbert space of quantum mechanics is then accomplished and it is used to derive the Schr\"odinger and Dirac equations and the breaking of the commutativity of spacetime geometry. The formulation provides an explanation of how determinism and random statistical behavior coexist in spacetime and a framework is developed that allows dynamical processes to be formulated in terms of chains of digits. These methods lead to a new (pre-geometrical) foundation for Lorentz transformations and special relativity. The Parker-Rhodes combinatorial hierarchy is encompassed within our approach and this leads to an estimate of the interaction strength of the electromagnetic and gravitational forces that agrees with the experimental values to an error of less than one thousandth. Finally, it is shown how the principle of least-radix economy naturally gives rise to Boltzmann's principle of classical statistical thermodynamics. A new expression for a general (path-dependent) nonequilibrium entropy is proposed satisfying the Second Law of Thermodynamics. 
\end{abstract}

%\begin{acknowledgements}
%\end{acknowledgements}



%\documentclass[prl,twocolumn,usletter,showpacs,superscriptaddress]{revtex4}
%%\documentclass[pre,preprint,usletter,showpacs,superscriptaddress]{revtex4}
%
%\usepackage{amsmath}
%\usepackage{amsfonts}
%\usepackage{amssymb}
%
%\usepackage [ansinew] {inputenc}
%%\usepackage[pdftex]{graphicx}
%\usepackage[dvips]{graphicx}
%%\usepackage[super, comma, sort&compress]{natbib}
%\usepackage{natbib}
%
%\begin{document}
%\title{Quantum Mechanics and the Principle of Least Radix Economy}
%
%\author{Vladimir Garc\'{\i}a-Morales}
%
%\affiliation{Institute for Advanced Study - Technische Universit\"{a}t M\"{u}nchen, Lichtenbergstr. 2a, D-85748 Garching, Germany}
%\affiliation{Nonequilibrium Chemical Physics - Physics Department - Technische Universit\"{a}t M\"{u}nchen, James-Franck-Str. 1, D-85748 Garching, Germany}
%
%\begin{abstract}
%\noindent 
\section{Introduction}

It has been conjectured that all natural processes can be understood as the result of computation \cite{Wolfram} \cite{Fredkin} \cite{McCauley}. This statement is contained in Wolfram's \emph{principle of computational equivalence} \cite{Wolfram} which is closely related to the Church-Turing thesis \cite{Turing}: A computable function (expressing e.g. a law of physics) is also  effectively calculable (i.e. its values can be found by some purely mechanical process). This thesis is the base for digital physics, in which the universe is modeled as a giant computer \cite{Wolfram} \cite{Fredkin} \cite{Hooft} processing the information contained in it. Ideas of digital physics had been independently advanced in the 60's in the context of quantum mechanics by Bastin et al. \cite{Bastin} \cite{Parker}, and then by Noyes and Kauffman \cite{Noyes} \cite{Kauffman1} \cite{Kauffman2}. In the context of nonlinear dynamics, McCauley and Palmore \cite{McCauley} \cite{MC2} \cite{MCPAL1} showed how converting real numbers into finite strings of digits from a finite alphabet can capture all dynamical features of chaotic deterministic systems. Very recently, these latter ideas have been systematically explored \cite{comphys} by means of $\mathcal{B}$-calculus \cite{VGM1}, which constitutes a mathematical formalism for rule-based dynamical systems (examples being cellular automata \cite{VGM1} \cite{VGM2} \cite{VGM3} and substitution systems \cite{VGM4}).

If one accepts something like the principle of computational equivalence described above, several questions can  be raised. First of all, since computations are assumed to be made with symbols of a finite alphabet, one can ask \emph{what the size (cardinal) $\eta$ of the alphabet should be and what physical meaning is to be attributed to the alphabet's size} (note that in this view computations are intended to directly map physical processes). In performing computations, the size of the alphabet coincides with the radix (base) in which a number is expressed. We shall henceforth use the latin word \emph{radix} as a synonym for ``base'' to avoid confusion with other uses of the latter word in physics. Because of its obvious, useful connection with boolean algebra, the binary radix has long been considered in the research of discrete physics \cite{Wolfram} \cite{Noyes} \cite{McCauley}.

In this article we present a new approach to quantum mechanics inspired by digital physics which gives an answer to the above questions. The non-commutativity of the continuum spacetime at the quantum level is \emph{derived} from our approach. We claim that \emph{nature makes dynamically the most effective choice for the radix in which its computations take place}. It is then shown that classical and quantum physics merge together from a single variational principle. An integer function of the dimensionless Lagrangian action $\lfloor S/h \rfloor$ (here $\left \lfloor x \right \rfloor$ denotes the floor function (lower closest integer) of $x$, $S$ is the Lagrangian action and $h$ is the Planck's constant) is interpreted as the radix in which the computations that implement the laws of nature take place. \emph{By demanding that this radix works most efficiently physical laws are derived in a unified way}. The Lagrangian action can thus be understood as a key quantity for the effectiveness of mathematics in the natural sciences \cite{Wigner}.

The outline of this article is as follows. In Section \ref{radie} the central idea of the article, involving  a quantity called radix economy (or digit capacity), is presented and illustrated with examples. The mathematical (and technological) relevance of this quantity (and the natural necessity of having it ``economical'') is substantiated in an attempt to provide an easy access to the later developments in the article (which explore its physical relevance). In Section \ref{PLRE} the principle of least radix economy is presented in detail. Three postulates in which our whole approach is based are stated. We then show how the principle gives rise to two kinds of physical solutions: least action paths and complex wavefunctions. The Hilbert space of quantum mechanics is then systematically constructed. This is accomplished through the specification of a complete orthonormal base and an inner product that are naturally linked to the structure of the solutions of the principle of least radix economy. These are classified into symmetry classes through cyclic groups described by the automorphisms that leave invariant each vector of the complete base of the Hilbert space. In Section \ref{Schro} the Schr\"odinger equation is derived, the eikonal approximation and the correspondence principle are proved, and our approach is compared to Feynman's path integral formulation (which is also based on the physical, Lagrangian action), briefly discussing similarities and differences (which are not of physical but of mathematical character). In Section \ref{noncom} we elucidate how the commutativity of spacetime is broken in the quantum realm and we illustrate how the main quantum numbers emerge from the description without need of solving the Schr\"odinger equation. We also show how finite dimensional Hilbert spaces can be naturally accounted for. In Section \ref{statistics} the physical radix is related to the mode of certain binomial distributions, whose form is established and discussed. These concepts are then linked to the possibility of expressing physical dynamical processes as chains of zeroes and ones. The physical content and meaning of these chains is better understood in terms of the unary radix. In Section \ref{unary} the unary radix and the consequences of the third postulate of Section \ref{PLRE} are explored and the quantum of action is investigated. Surprisingly, the statistical distribution of the latter provides a key to understanding how Minkowski geometry arises locally in spacetime at a classical level. The quantum of action, together with the principle of least radix economy, allows to understand how different scales, described by different values of the optimal radix, are interrelated and how particles can be equivalent to fields. Special relativity is \emph{derived} from these insights and relativistic wave equations that describe particles with spin (Dirac equation) are also derived. In Section \ref{entropy} we show how classical statistical thermodynamics and the Second Law of Thermodynamics emerge from the principle of least radix economy. Finally, in Section \ref{Parker} we explain how our approach encompasses the Parker-Rhodes combinatorial hierarchy, giving a direct and accessible physical meaning to this intriguing mathematical concept and showing how it indeed provides an argument to establish the value of fundamental physical constants governing the strength of fundamental interactions.

\section{Radix economy and the Lagrangian action} \label{radie}

We usually distinguish the physical impact of numbers in terms of the orders of magnitude that they involve. A related (but so far unexplored) approach is to consider the radix in which numbers are expressed. The decimal radix which we always adopt for representing numbers in physics is just a tacit convention which is not necessarily the most efficient one. The \emph{representation} of all numbers with physical meaning is radix dependent. For example, if we consider a length of 123 meter, it is understood that we mean $1\cdot 10^{2}+2\cdot 10^{1}+3\cdot 10^{0}$ m, i.e that we give the number through its representation in the decimal radix. Had we used another radix, 50 say, this same representation would yield $1\cdot 50^{2}+2\cdot 50^{1}+3\cdot 50^{0}$ m and with the same figures we would mean another number, i.e. 2603 m, instead. We all conventionally (and tacitly) agree to represent all our numbers in the same fixed radix so that we can easily compare them (if we fix the representation instead, we see that by changing the radix we change, e.g., a length). Let $b$ (a natural number) denote an arbitrary radix. We can represent any real number $A$ in radix $b$ as 
\begin{equation}
A= \sum_{m=-\infty}^{\left \lfloor 1+\log_{b}A\right \rfloor}b^{m-1}\mathbf{d}_{b}(m, A)  \label{fund1}
\end{equation} 
where the upper bound in the sum $\left \lfloor \log_{b}A+1\right \rfloor$ is the total number of integer digits of $A$ in radix $b$ \cite{comphys}, and $\mathbf{d}_{b}(m, A)$ is an integer function that returns the digit of $A$ that accompanies the $(m-1)$th-power of $b$ when $A$ is written in base $b$. This latter function \cite{comphys} (see also \cite{Knuth}), yields an integer between 0 and $b-1$ and is defined as  
\begin{equation}
\mathbf{d}_{b}(m, A) \equiv \left \lfloor \frac{A}{b^{m-1}} \right \rfloor - b \left \lfloor \frac{A}{b^{m}}\right \rfloor \label{digit}
\end{equation}

An important means to quantify the effectiveness of radix $b$ to express the number $A$ is the \emph{radix economy} $\mathcal{C}(b,A)$, also called \emph{digit capacity} \cite{Hurst} 
\begin{equation}
\mathcal{C}(b,A)=b \left \lfloor 1+\log_{b}A\right \rfloor \label{dcap}
\end{equation}
This quantity is related to the hardware economy in circuits with multiple valued logic \cite{Hurst}.
When it is a minimum, we say that radix $b$ expresses most efficiently $A$ or that $b$ has the least radix economy in expressing $A$. Why this quantity is related to hardware economy can be understood as follows: As mentioned above, if we want to give the integer number $A$ in base $b$ we need $\left \lfloor 1+ \log_{b}A\right \rfloor$ figures since the largest power of $b$ that is needed to express $A$ in radix $b$ is $\left \lfloor \log_{b}A\right \rfloor$ and 0 is the lowest power. Each of these figures can be any integer number $\in [0, b-1]$ which means that, in order to render an arbitrary number of the same order of magnitude as $A$ in base $b$ each position should be able to accommodate any of the $b$ figures. In a digital device, if one then assumes that one needs, e.g. a number $b$ of light-emitting diodes per figure (a subset of them being illuminated to render any figure between $0$ and $b-1$) this then means a ``hardware cost'' precisely given by Eq. (\ref{dcap}), i.e. $b$ times the total number of digits of the number to be represented. In the examples above the representation $123$ has three digits in both radices $b=10$ and $b=50$. Hence, the numbers associated to this representation have digit capacities $10 \cdot 3=30$ for the one in the decimal radix and $50\cdot 3=150$ for the one in radix $b=50$. Therefore, the decimal radix is more economic in this case. 

Another example to illustrate the concept of radix economy is the following. Suppose that you want to design a pocket calculator to work with numbers whose order of magnitude would never exceed $10^{6}$ and you envisage to have up until 10 figures of space to represent numbers. Would the decimal radix be the most economic choice in this case? To answer this question, note that the number $10^{7}-1=9999999$ needs seven figures to be represented in the decimal radix. The radix economy would thus be $\mathcal{C}(10,10^{7}-1)=10 \left \lfloor 1+\log_{10}(10^{7}-1)\right \rfloor=10\cdot 7=70$. However, a more economic radix would be 6, since $10^{7}-1$ is represented as 554200143 in radix 6, i.e. a number with 9 figures, and hence $\mathcal{C}(6,10^{7}-1)=6 \left \lfloor 1+\log_{6}(10^{7}-1)\right \rfloor=6\cdot 9=54$, which proves to be more economic than the decimal radix: numbers of 9 figures in radix 6 are more economic than numbers of 7 figures in the decimal radix!

From Eq. (\ref{fund1}) we have
\begin{eqnarray}
A&=&\sum_{m=1}^{\lfloor \log_{b}A \rfloor+1} b^{m-1} \mathbf{d}_{b}(m,A) =\sum_{k=0}^{b-1}\sum_{m=1}^{\lfloor \log_{b}A \rfloor+1} kb^{m-1} \delta_{k,\mathbf{d}_{b}(m,A)}
\label{idenA}
\end{eqnarray}
where $\delta_{k,\mathbf{d}_{b}(m,A)}$ denotes the Kronecker delta (i.e. it returns one if $k=\mathbf{d}_{b}(m,A)$ and zero otherwise). \emph{The digit capacity, Eq. (\ref{dcap}), corresponds to the size of the matrix with elements $R_{km}$ given by}
\begin{equation}
R_{km}=\delta_{k,\mathbf{d}_{b}(m,A)} \label{matrep}
\end{equation} 
which is the \emph{representation matrix} of the integer $A$. The latter is generally rectangular and contains only zeros and ones. It has $b$ rows labelled by index $k \in [0,b-1]$ and $\lfloor \log_{b}A \rfloor+1$ columns, labelled by index $m$. The matrix has thus size given by Eq. (\ref{dcap}). For example, the number $A=11$ reads '102' in radix 3 and, in that radix, it has matrix representation
$
\scriptsize{\left( \begin{array}{rrr} 0 & 1 & 0 \\ 0 & 0 & 1 \\ 1 & 0 & 0 \end{array} \right)}
$. 
Its size is $9$, which coincides with $\mathcal{C}(3,11)$ given by Eq. (\ref{dcap}).
Thus, Eq. (\ref{idenA}) can be written in matrix form, which in this case reads 
\begin{eqnarray}
\left(0\quad 1\quad 2\right) \left( \begin{array}{rrr} 0 & 1 & 0 \\ 0 & 0 & 1 \\ 1 & 0 & 0 \end{array} \right)\left( \begin{array}{r} 3^{0} \\ 3^{1} \\ 3^{2} \end{array} \right)=
\left(2\quad 0\quad 1\right)\left( \begin{array}{r} 3^{0} \\ 3^{1} \\ 3^{2} \end{array} \right)=11
\end{eqnarray}
The left and right vectors are fixed. They have components $0, 1, \ldots, b-1$ and $b^{0}, b^{1}, \ldots, b^{\lfloor \log_{b}A \rfloor}$ respectively. Therefore, \emph{the matrix representation characterizes each number in each radix in a unique way}.

Quite interestingly, for $b=\left \lfloor A \right \rfloor >1$ we have
\begin{eqnarray}
\mathcal{C}(\left \lfloor A \right \rfloor,A)&=&\left \lfloor A \right \rfloor \left \lfloor 1+\log_{\left \lfloor A \right \rfloor}A \right \rfloor \nonumber \\
&=&\left \lfloor A \right \rfloor \left \lfloor 1+\frac{\ln A}{\ln \left \lfloor A\right \rfloor} \right \rfloor=2\left \lfloor A \right \rfloor \label{mag}
\end{eqnarray}
since we have $A=\left \lfloor A\right \rfloor +\epsilon$ with $0 \le \epsilon <1$ and thus
\begin{equation}
1 \le \frac{\ln A}{\ln \left \lfloor A\right \rfloor}=\frac{\ln (\left \lfloor A\right \rfloor +\epsilon)}{\ln \left \lfloor A\right \rfloor}\approx 1+\frac{\epsilon}{\left \lfloor A\right \rfloor \ln \left \lfloor A\right \rfloor} < 2
\end{equation} 
 
The unary radix $b=1$ does not obey Eq. (\ref{mag}) although it defines the most elementary numeral system (and the lowest possible bound for a numeral system with a natural base) and certainly has a useful physical meaning (as we shall see). This system is a most ancient one. It appears already in the Golenishchev Mathematical Papyrus, an ancient Egyptian papyrus which was most likely written down in the 13th dynasty of Egypt (roughly 1850 BC). Since zero does not exist in the unary radix, it is not possible to represent with it non-existing things. This probably explains why the ancient Egyptians knew a formula for the volume of a frustum (i.e. an unfinished pyramid) but not for the one of a finished pyramid: They did not have the number zero! (This idea is fully elaborated in \cite{Schroeder} p. 6.) 

In order to represent a natural number $A$ in the unary radix $b=1$, an arbitrarily chosen symbol '1' is repeated $A$ times. This numeral system, thus, transforms any number in the \emph{number of counts} in which '1' appears. Addition and subtraction in this radix amounts to concatenate/remove strings of 1's. The following examples show how numbers in the decimal radix are written in unary:
\begin{equation}
2=11 \qquad 3=111 \qquad 5=11111 \qquad 10=1111111111 \label{examples}
\end{equation}
Note that the position of the 1's is immaterial: all 1's are indistinguishable (it does not make sense to talk here about powers of the radix). Only their total number is important. Thus, the 1's behave just as tally marks.  The radix capacity $\mathcal{C}(1,A)$ in the unary radix is, thus, trivially 
\begin{equation}
\mathcal{C}(1,A)=A \label{raduno}
\end{equation}
since $A$ has $A$ digits in radix 1 and the alphabet has only one symbol. Thus, the unary radix is the one with the least economy to represent natural numbers $A \le 5$ (note that 6, which is written as 110 in the binary radix, has digit capacity $\mathcal{C}(2,6)=2\cdot 3=6$ in that radix and, therefore, it is equal to the capacity $\mathcal{C}(1,6)=6$ in the unary radix). 

We finish these remarks on numeral systems with the trivial case $b=0$ which does not constitute any useful numeral system, but which is, nonetheless, of physical relevance to represent the vacuum. In such radix, there are no natural numbers $A$ to be represented others than $A=0$. The radix economy is thus $\mathcal{C}(0,0)=0$. Such ``radix''  has always the least economy since, trivially, it does not cost anything to represent nothing. 

We have seen above that if a number $A_{1}$ has the same number of digits in radix $b_{1}$ as it has $A_{2}$ in radix $b_{2}$, then, if $b_{1} < b_{2}$ so is also $A_{1}<A_{2}$. We note also that a real number $A \ge 2$ in radix $\left \lfloor A \right \rfloor$ has always 10 as integer part since $\left \lfloor A \right \rfloor \le A < \left \lfloor A \right \rfloor+1$ and, therefore, $A=1\cdot \left \lfloor A \right \rfloor^{1}+0\cdot \left \lfloor A \right \rfloor^{0}+\{A\}$, where $\{A\}$ denotes the fractional part of $A$. Thus, crazy as it may seem, 10 m can denote any arbitrarily large length if we tune the radix in which 10 is expressed (note, however, that 1 m is always 1 m regardless of the radix used!).

We read in Dirac's book \cite{Dirac} (p. 3): \emph{So long as big and small are relative concepts, it is no help explaining the big in terms of the small. It is therefore necessary to modify classical ideas in such a way so as to give an absolute meaning to size.} We now show how the dimensionless Lagrangian action $S/h$, where $h$ is the Planck constant, can be used to that purpose. Along a path $\gamma$ connecting two points '1' and '2', $S$ is an scalar functional $\mathbb{R}^{N} \to \mathbb{R}$ given by 
\begin{equation}
S(\mathbf{q}(t)) = \int_{t_1}^{t_2} L(\mathbf{q}(t),\dot{\mathbf{q}}(t),t)\, dt 
\end{equation}
where $L$ is the Lagrangian, $t$ denotes time, and $\mathbf{q}(t)$ and $\dot{\mathbf{q}}(t)$ are the generalized position and velocity vectors evaluated along the path. When $S/h$ is large the physical trajectories are governed by the principle of least action
\begin{equation}
D_{\varepsilon}S(\mathbf{q}(t))  \equiv  \left.\frac{d}{d\varepsilon} S(\mathbf{q}(t) + \varepsilon f)\right|_{\varepsilon = 0}=0 \label{prinA}
\end{equation}
where $D_{\varepsilon}$ denotes the first-variation operator, $\varepsilon$ is a scalar and $f$ is an arbitrary function. The extremization of the action leads to the Euler-Lagrange equations describing the physical trajectories
\begin{equation}
\frac{d}{dt} \frac{\partial L}{\partial \dot q_{i} }-\frac{\partial L}{\partial q_{i}}=0 \qquad i \in [1,N] \label{Euler}
\end{equation} 
where $N$ is the total number of degrees of freedom in the system and $q_{i}$ and $\dot{q}_{i}$ are the $i$-th component of the generalized position and velocity vectors, respectively. An equivalent formulation of the equations of motion is provided by the Hamilton-Jacobi equation 
\begin{equation}
H\left(\mathbf{q}, \nabla S; t\right)+\frac{\partial S}{\partial t}=0 \label{HJE}
\end{equation}
which implies the following relationships involving the Hamiltonian $H$ and generalized momenta $\mathbf{p}$ as 
\begin{equation}
H = - {\partial S \over \partial t} \qquad \ \ \mathbf{p}=\nabla S  \label{HJ}
\end{equation}
When $S/h$ is small, there are no longer well defined unique trajectories along which the motion of the system takes place. Rather, infinitely many choices are possible and one is forced to speak about the probability of finding a certain physical state. This is a most striking fact in quantum mechanics: When the action $S$ is of the order of $h$ the physical laws seem very different to when $S/h$ is large. Thus, the (generally real) number $S/h$ suggests a way to give an absolute meaning to size, breaking the relativity of big and small.

The question thus arises as whether we can extract a physical radix from the real-valued quantity $S/h$ which is to be considered as the characteristic ``physical length'' of a certain dynamical process. If we introduce the natural demand that the radix (an integer) should increase \emph{linearly} on the integers as does $S/h$ on the reals, a straightforward way to achieve this is to take $\lfloor S/h \rfloor$ as the physical radix by observing that
\begin{equation}
\frac{S}{h}=\left \lfloor \frac{S}{h} \right \rfloor+\left \{ \frac{S}{h} \right \} \label{zerteilt}
\end{equation}
i.e. that $S/h$ is the sum of its integer and its fractional parts. The r.h.s. is the sum of two nonlinear discontinuous functions on the reals, the integer-valued function $\left \lfloor \frac{S}{h} \right \rfloor$ which is a \emph{staircase} with discrete jumps at each integer value of $S/h$ (and which strongly approaches $S/h$ for $S$ large, being always \emph{linear} on the integers), and the real-valued function $\left \{ \frac{S}{h} \right \}$, which is a \emph{periodic} function of $S/h$ (taking values between $0$ and $1$) with the form of a sawtooth wave, being discontinuous at the integer values of $S/h$ and \emph{linear} everywhere else.

The principle of least action only works \emph{asymptotically}, when $S/h$ is  large. In such limit $\left \{ \frac{S}{h} \right \}$ can be neglected and we have, as already mentioned $S/h \approx \left \lfloor S/h \right \rfloor$. Therefore, from Eq. (\ref{zerteilt}), we obtain
\begin{equation}
\frac{1}{h}D_{\varepsilon}S \sim D_{\varepsilon}\left \lfloor \frac{S}{h} \right \rfloor=\frac{1}{2}D_{\varepsilon}\mathcal{C}\left(\left \lfloor \frac{S}{h} \right \rfloor, \frac{S}{h} \right) \label{atten}
 \end{equation}
where we have used Eq. (\ref{mag}) in getting to the last equality. This, therefore, leads to reinterpret the \emph{least (classical) action paths} as being \emph{those for which the radix $\left \lfloor S/h \right \rfloor$ is most efficient}, i.e. \emph{those paths for which the radix $\lfloor S/h \rfloor$ has the least economy}. With ``efficiency'' we mean those laws which extremize the digit capacity, Eq. (\ref{dcap}), with $b \equiv \lfloor S/h \rfloor$ and $A=S/h$. As we have seen above this efficiency represents an \emph{actual mathematical efficiency} since it also means that the matrix representations of physical numbers, Eq. (\ref{matrep}), have the \emph{lowest} size. We propose that this principle of least radix economy holds \emph{generally} (also in situations where $S/h$ is not necessarily large) thus providing foundations for classical and quantum mechanics and statistical thermodynamics as well. In the next sections we substantiate this claim. 

Note that $\left \lfloor S/h \right \rfloor$ is an \emph{integer-valued discontinuous function} and the first variation operator,  infinitesimally acts on well-behaved (real-valued) functionals. From a mathematical point of view this would seem a major violence (belonging to the kind of problems mentioned in \cite{Neumann}, p. 28). However, by noting that $\left \lfloor \frac{S}{h} \right \rfloor=\frac{S}{h}-\left \{ \frac{S}{h} \right \}$ and $\left \{ \frac{S}{h} \right \}$ is a periodic function we have \cite{Tits}
\begin{equation}
\left \{ \frac{S}{h} \right \}=\frac{1}{2} - \frac{1}{\pi} \sum_{k=1}^\infty \frac{\sin \left(2 \pi k \frac{S}{h}\right)}{k} \label{saw}
\end{equation}
We claim that this Fourier series, which converges to $\{ S/h\}$ everywhere when $S/h$ is noninteger, is a \emph{physically meaningful} representation of $\{ S/h\}$ even at the discontinuities $S/h$ integer (where Gibbs phenomenon takes place \cite{Dym}). 

Together with the decomposition of Eq. (\ref{zerteilt}) a mathematical result that we shall frequently use is the following. We note that
\begin{equation}
\eta \le S/h < \eta+1
\end{equation} 
and therefore
\begin{equation}
0 \le \frac{\eta}{\eta+1} \le \frac{S/h}{\eta+1} < 1 \label{desigul}
\end{equation} 
This means that $\frac{S/h}{\eta+1}$ is equal to its fractional part $\left \{ \frac{S/h}{\eta+1} \right \}$. Thus 
\begin{equation}
\frac{S}{h}= (\eta+1)\left \{\frac{S/h}{\eta+1} \right \}. \label{eqmode1}
\end{equation}
and we also have, therefore,
\begin{equation}
\eta=\left \lfloor \frac{S}{h} \right \rfloor=\left \lfloor (\eta+1)\frac{S}{(\eta+1)h} \right \rfloor=\left \lfloor (\eta+1)\left \{\frac{S}{(\eta+1)h}\right\} \right \rfloor \label{eqmode}
\end{equation}

\section{The principle of least radix economy} \label{PLRE}

We have thus seen that a fixed radix is not always the most efficient choice in all situations involving representations of numbers (e.g. the example of the pocket calculator above). Of course, \emph{in any specific scale the description of physical laws is not affected in any way by the choice of the radix since the latter is merely an (equivalent) representation of the numbers}. 
Since Physics is not scale invariant (from the considerations made in the previous section on the role of $S/h$) there is, however, a fundamental physical radix: \emph{the one which most efficiently works at each scale}. Furthermore, \emph{the fundamental radix establishes the form that physical laws do have at each scale}. We propose that nature has her own \emph{dynamical} means to specify the most efficient radix. Using action as our guide, we thus introduce the following postulates.

~\\
\noindent\fbox{%
    \parbox{\textwidth}{
    
\noindent \textbf{A}. The physical radix (a natural dimensionless number) is given by 
\begin{equation} 
\eta \equiv \left \lfloor \frac{S}{h} \right \rfloor=\left\{
\begin{array}{ll} \frac{S}{h} - \frac{1}{2} + \frac{1}{\pi} \sum_{k=1}^\infty \frac{\sin \left(2 \pi k \frac{S}{h}\right)}{k} & \qquad \text{if } \frac{S}{h} \notin \mathbb{Z} \\ \tiny{} &\small{} \\
\frac{S}{h} & \qquad \text{if } \frac{S}{h} \in \mathbb{Z}
\end{array} \right. \label{etadef}
\end{equation}

\noindent \textbf{B}. (\emph{Principle of least radix economy}) Let $\mathcal{A}$ be either (a) an action-like radix-dependent functional or (b) a dimensionless radix-independent quantity arising from, e.g., a counting argument. Then, physical laws are derived by extremizing the radix economy
\begin{equation}
D_{\varepsilon}\mathcal{C}\left(\eta, \mathcal{A} \right)=D_{\varepsilon} \left( \eta \left \lfloor1+\log_{\eta} \mathcal{A} \right \rfloor \right)=0  \label{capac}
\end{equation}
with $\eta >1$ given by postulate \textbf{A}. In case (a) $\mathcal{A}$ is simply called \emph{action} and we shall take $\mathcal{A}=S/h$. In case (b) $\mathcal{A}$ is called a \emph{(generalized) partition function} and the least radix economy $\mathcal{C}$ is said to be an \emph{entropy} or a \emph{(generalized) Massieu-Planck potential}.

\noindent \textbf{C}. The unary radix $\eta=1 \equiv \eta_{1}$ gives the quantum of action. When it has the least economy, it describes \emph{particles} if $S/h=\eta_{1}$ (hence $\{S/h\}= 0$) and \emph{fields} (the quantum smeared in spacetime) otherwise. $\eta=0 \equiv \eta_{0}$ describes the vacuum and has always the least economy. Therefore, the vacuum is always present in any physical situation.

    }%
}
\medskip

We shall discuss postulate \textbf{C} in Section \ref{unary} and case (b) of postulate $\textbf{B}$ in Section \ref{entropy}.  We now focus on case (a) of postulate $\textbf{B}$ with $\eta \ge 2$. Thus, with $\mathcal{A}=S/h$ we have, from Eqs. (\ref{mag}) and (\ref{etadef})
%\begin{equation}
$\mathcal{C}\left(\eta, S/h \right)=\eta \left \lfloor \log_{\eta}\left(S/h\right)+1\right \rfloor=2\eta=2\left \lfloor S /h \right \rfloor  
$%\end{equation}
and hence, the principle of least radix economy takes the form 
\begin{equation}
D_{\varepsilon}\mathcal{C}\left(\eta,S/h \right)=2 D_{\varepsilon}\left \lfloor S /h \right \rfloor= 0 \label{prin}
\end{equation}
Therefore
\begin{eqnarray}
%D_{\varepsilon}\mathcal{C}&=&2D_{\varepsilon} \eta=
2D_{\varepsilon}\left \lfloor S /h \right \rfloor  &=&
2D_{\varepsilon}\left(\frac{S}{h} - \frac{1}{2} + \frac{1}{\pi} \sum_{k=1}^\infty \frac{\sin \left(2 \pi k \frac{S}{h}\right)}{k}\right) \qquad \qquad \nonumber \\
&=&\frac{2}{h}\left[1+2\sum_{k=1}^{\infty}\cos\left(\frac{2\pi kS}{h}\right)\right]D_{\varepsilon} S \nonumber \\
&=& \frac{2}{h}\sum_{k=-\infty}^{\infty}e^{i2\pi kS/h}D_{\varepsilon} S  = \frac{2}{h}\mathcal{D}_{\infty}(2\pi S/h)D_{\varepsilon} S \nonumber \\
&=& \frac{4\pi}{h}\sum_{n=-\infty}^{\infty}\delta\left(\frac{2\pi S}{h}-2\pi n\right)D_{\varepsilon} S \nonumber \\
&=& \frac{2}{h}\delta\left(\left \{ \frac{S}{h}\right \} \right)D_{\varepsilon} S =0
\label{vari}
\end{eqnarray}
where $\mathcal{D}_{\infty}(x)\equiv \sum_{k=-\infty}^{\infty}e^{ikx}$ is the Dirichlet kernel and $\delta(x)$ is the Dirac delta function \cite{Lanczos}. There are thus two kinds of possible physical solutions: 

\begin{itemize}
\item I. Paths where the  action $S$ is minimized (i.e. for which $D_{\varepsilon} S=0$); 

\item II. \emph{Any} path where the action $S$ is \emph{not} an integer multiple of $h$  (since then the Dirac comb in Eq. (\ref{vari}) is zero). These are relevant in the quantum regime which, from Eq. (\ref{etadef}) takes place when 
\begin{equation}
\frac{S}{h} \sim \frac{1}{2} - \frac{1}{\pi} \sum_{k=1}^\infty \frac{\sin \left(2 \pi k \frac{S}{h}\right)}{k} = \left \{ \frac{S}{h} \right \} \label{crit}
\end{equation}
\end{itemize}

When Eq. (\ref{crit}) holds, paths \emph{where $S$ is an integer multiple of $h$} are also physically meaningful only if they are of the Type I as well, i.e. \emph{if they are least action paths satisfying Eq. (\ref{prinA})}. These paths reproduce indeed the Bohr quantization rule where the action is first minimized and then a value $nh$ is attributed to the least action trajectories. All these considerations guarantee that the classical limit governed by Eq. (\ref{prinA}) is asymptotically approached from Eq. (\ref{vari}) for $S/h$ large through the semiclassical quantization rules, as was also already observed in the matrix formulation of quantum mechanics \cite{Heisenberg}: In this limit $S/h \sim \left \lfloor S/h \right \rfloor$, Eq.(\ref{crit}) does not hold ($\{ S/h \}$ can be neglected compared to $S/h$) and Eq. (\ref{vari}) reduces to the least action principle, Eq. (\ref{prinA}). This is the correspondence principle, which we shall rigorously prove in Section \ref{Schro}.

Henceforth, \emph{we shall always assume that $S/h$ can take any arbitrary value having in mind that when it is integer it corresponds to a least-action path as well}. Any solution of the Euler-Lagrange differential equations, Eq. (\ref{Euler}), is automatically also a solution of Eq. (\ref{vari}). Let us look for other solutions. First we note a crucial fact of Eqs. (\ref{zerteilt}), (\ref{etadef}) and (\ref{vari}): they are all invariant under the discrete transformation
\begin{equation}
S \to S+mh \label{sime}
\end{equation}
with $m$ integer.  Hence, if $S$ is a solution of the variational principle so must be $S+mh$ necessarily as well. We now prove the following result: \emph{If $S/h$ is irrational (and, hence, a solution of Type \emph{II}) the (multivalued) inverse function $\frac{S'}{h} \equiv \frac{1}{i2\pi k}\ln e_{k} $ of any member of the family}
\begin{equation}
e_{k}=e^{ik2\pi S/h} \qquad \qquad k=0, \pm 1, \pm 2, \ldots, \pm \infty \label{ortonor}
\end{equation}
\emph{is a solution of Type \emph{II} of Eq. (\ref{vari}) as well.} 

To prove this we must only check that Eq. (\ref{vari}) is satisfied for $S'/h$, i.e., that we have
\begin{equation}
 \frac{4\pi}{h}\sum_{n=-\infty}^{\infty}\delta\left(\frac{2\pi S'}{h}-2\pi n\right)D_{\varepsilon} S' =0 \nonumber
\end{equation}
We obtain, 
\begin{eqnarray}
&&  \frac{4\pi}{h}\sum_{n=-\infty}^{\infty}\delta\left(\frac{2\pi S'}{h}-2\pi n\right)D_{\varepsilon} S' \nonumber \\
&&=\frac{4\pi}{h}\sum_{n=-\infty}^{\infty}\delta\left(\frac{1}{ik}\ln e_{k}-2\pi n\right)D_{\varepsilon}\left( \frac{h}{i2\pi k}\ln e_{k}\right)
\nonumber \\
&&=\frac{4\pi}{h}\sum_{n=-\infty}^{\infty}\delta\left(\frac{1}{ik}\left[ \ln 1+i(\arg(e_{k})+2\pi q)\right]-2\pi n\right)D_{\varepsilon} \left( \frac{h}{i2\pi k}\ln e_{k}\right) \nonumber \\
&&=\frac{4\pi}{h}\sum_{n=-\infty}^{\infty}\delta\left(\frac{\arg(e_{k})}{k}+\frac{2\pi q}{k}-2\pi n\right)D_{\varepsilon} \left( \frac{\arg(e_{k})}{2\pi k}h+\frac{q}{k}h \right) \nonumber \\
&&=\frac{4\pi}{h}\sum_{n=-\infty}^{\infty}\delta\left(\frac{2\pi S}{h}+\frac{2\pi q}{k}-2\pi n\right)D_{\varepsilon} \left(S+\frac{q}{k}h\right) \nonumber \\
&&=\frac{4\pi}{h}\sum_{n=-\infty}^{\infty}\delta\left(\frac{2\pi S_{q/k}}{h}-2\pi n\right)D_{\varepsilon}S_{q/k}=0 \label{Sq}
\label{vari2}
\end{eqnarray}
where $q$ is an arbitrary integer and we have defined $\frac{S_{q/k}}{h}\equiv \frac{S}{h}+\frac{q}{k}$. In getting to the last equality we have used the fact that, since $S/h$ is irrational so is $\frac{S_{q/k}}{h}$ and, hence, the Dirac comb is zero because it is not possible to have $2\pi S_{q}/h=2\pi n$ for any integers $n, q$ and $k$. This proves the result. Since the irrational numbers are the most abundant ones in the real line (the rational numbers having measure zero) \cite{Niven} this result is expected to be most important in the deep quantum regime (where irrational numbers should play a most prominent role) and we shall use it below.

The multivaluedness of the complex logarithm gives thus rise to symmetric partners of $S$ called $S_{q/k}$ ($S_{0}=S$) which correspond to complex numbers in the circle $S^{1}$ given by Eq. (\ref{ortonor}). Therefore, together with values $S_{mk/k}=S_{m}=S_{0}=S$ on different sheets of the Riemann surface [which are symmetric by virtue of Eq. (\ref{sime})] there appear $k$ symmetric partners $S_{q/k}$ in each sheet of the Riemann surface as well, which correspond to the action values $S_{0/k},\ S_{1/k},\ S_{2/k},\ \ldots, \ S_{(k-1)/k}$. They are induced in the circle $S^{1}$ through the action of the group of automorphisms $x \to 2\pi q x/k$ with $q=0, 1, \ldots , k-1$ which leave $e_{k}$ invariant under composition. This group is isomorphic to the finite cyclic group $\mathbb{Z}_{k}=\mathbb{Z}/k\mathbb{Z}$.  

Two interesting corollaries are easily derived from the above result. \emph{If we set $k=1$ in Eq. (\ref{Sq}), then if $S/h$ is a solution of the variational principle $\frac{1}{i2\pi}\ln e_{1}(S/h)$ is also a solution of the same type}. That this is so comes from the discrete symmetry Eq. (\ref{sime}). As we shall prove below (see Section \ref{Schro}), this corollary is directly connected to the eikonal approximation in the semiclassical regime. 

The second corollary is that \emph{if $S/h$ is a rational solution, there generally exist physical paths with $S_{q/k}/h$ integer and $k>1$ which necessarily correspond to least-action trajectories}.  This result may be helpful in understanding subtleties of the semiclassical regime which proved to be stumbling blocks for the old quantum theory. For example, it provides an explanation of why the Heisenberg semiclassical quantization of the Helium atom with half-integer quantum numbers seemed to work well in certain cases. Because of symmetry considerations, the quantum physical state of the Helium atom, having two electrons, appears to be naturally described by the member $e_{2}=e^{i4\pi S/h}$, which induces solutions $S_{0/2}, S_{1/2}$ with a $\mathbb{Z}_{2}$ symmetry in $S^{1}$ [although we must refrain from giving a detailed calculation here, that this should be the case is because this symmetry group naturally reflects the exchange degeneracy of the two electrons (see e.g. \cite{Bohm}, p. 480)]. Because of the symmetry Eq.(\ref{sime}), there exist states with $S_{1/2}/h+m$ integer which correspond to half-integer values for the action $S/h=S_{0/2}/h$. Therefore, from the corollary, \emph{since $S/h$ takes half-integer values, this gives rise to semi-classically quantized least-action periodic paths as the ones calculated by Heisenberg}. Since the Helium atom constitutes a three-body problem \cite{Tanner} in spite of the strong correlations induced by $e_{2}$, all other members $e_{k}$ should also be present leading to quantum chaos. Gutzwiller's trace formula \cite{Gutzwiller} has been found to provide an excellent account of quantum chaos in the semiclassical regime. In the semiclassical regime, all our above arguments are fully consistent with the importance that classical periodic orbits play in Gutzwiller's theory \cite{Gutzwiller}. Here, such periodic orbits are ascribed to the Riemann surfaces created by Eq. (\ref{sime}) through the members $e_{k}$ and the action values $S_{q/k}/h$ in each sheet of the Riemann surface. 
 
The main \emph{consequence} of the above theorem and corollaries is the following: \emph{there exists a direct physical correspondence between the action $S/\hbar$ on a physical path and the phase $\chi$ of a complex number $e^{i\chi}$ in the unit circle $S^{1}$.}  The reverse statement is then also true, albeit in a different form: \emph{any point $e^{i\chi}$ in the unit circle $S^{1}$ corresponds to a certain (possibly infinite) ensemble of paths in physical space}. Any linear combination $\psi(S)$ of the mappings $e_{k}(x): \mathbb{R} \to S^{1}$ in the complex plane yielding the trigonometric series (or polynomial) \cite{Zygmund}
\begin{equation}
\psi(S)=\sum_{k=-\infty}^{\infty}\widetilde{\psi}(k)e^{i2\pi kS/h}\equiv e^{i\chi} \label{genform}
\end{equation}
is \emph{surjective}. Indeed, this is related to the fact that $\mathbb{R}$ is the universal cover of $S^{1}$. Note that $\chi$ is in Eq. (\ref{genform}) a kind of ``averaged action'' which is itself a solution of the principle of least radix economy. The multivalued inverses of each $e_{k}(x)$ taken separately are complex logarithms whose values unfold the possible (real) values for the action with different discrete symmetries indexed by $k$. We have proved all the latter to be a solution of Eq. (\ref{vari}) by virtue of Eq. (\ref{sime}). Indeed, the mapping $e_{k}(x)$ can be understood as the composition of two mappings (the first being already surjective) $\{x\}: \mathbb{R}\to [0,1)$ and $[e_{k} \circ \{ \}]x=e_{k}(\{ x \}): [0,1) \to S^{1}$.

We have now tools to study the mathematical structure of solutions in physical space that do not necessarily correspond to least-action trajectories but which have a significant dynamical impact in the quantum realm. From the above correspondence we have found that \emph{any arbitrary sequence of complex numbers whose sum converges to a complex number in $S^{1}$ maps to a physical solution of the principle of least radix economy}. Therefore, since the trigonometric series Eq. (\ref{genform}) converges to a complex number of unit modulus $\psi(S) \in S^{1}$, it maps to physical solutions. Eq. (\ref{genform}) is called a \emph{wavefunction} of the physical system, with the property $\psi\psi^{*}=|\psi|^{2}=1$, i.e. the modulus of the complex number $\psi$ is unity (the asterisk denotes complex conjugation). By the Cauchy-Schwarz inequality \cite{Weyl} we then observe that
\begin{equation}
\sum_{k=-\infty}^\infty |\widetilde{\psi}(k)|^2 < \infty  \label{conve}
\end{equation}
i.e., the sum converges to a finite positive real value: the norm. Hence, the $\widetilde{\psi}(k)$'s are terms of the sequence space $\ell^{2}$ which consists of all convergent Cauchy sequences. This sequence space is a most prominent example of Hilbert space $\mathcal{H}$. In fact, it was through this example that Hilbert introduced his theory of linear integral equations \cite{Akhiezer}. There, thus, exist an scalar product 
\begin{eqnarray}
(\psi_{A} ,\psi_{B} )&=&\sum_{k'=-\infty}^\infty\sum_{k=-\infty}^\infty \int_{\eta}^{\eta+1} \widetilde{\psi}_{A}^{*}(k')\widetilde{\psi}_{B}(k)e^{-i2\pi (k'-k)\{S/h\}}d\left(S/h\right) \nonumber \\
&=&\sum_{k=-\infty}^\infty\sum_{k'=-\infty}^\infty \widetilde{\psi}_{A}^{*}(k')\widetilde{\psi}_{B}(k) \delta_{kk'}=
\sum_{k=-\infty}^\infty \widetilde{\psi}_{A}^{*}(k)\widetilde{\psi}_{B}(k)  \label{escalar}
\end{eqnarray}
which is a complex-valued function of two vectors $\psi_{A}, \psi_{B}$ $\in \mathcal{H}$ constituting a positive definite hermitian form. A scalar function $(f,g): X \times X \to \mathbb{C}$ (or $\mathbb{R}$ if $X$ is a real linear space) is said to be an inner product or scalar product if it satisfies the following conditions:
~\\

\noindent i. $(f,f)\ge 0 \quad \forall f \in X$ with equality if and only if $f=\textbf{0}$.

\noindent ii. $(f,g)=\overline{(g,f)} \quad \forall f, g \in X$. (The overline denotes complex conjugation.)

\noindent iii. $(af,g)=a(f,g) \quad \forall f, g \in X$ and $a \in \mathbb{C}$ (or $\mathbb{R}$).

\noindent iv. $(f_{1}+f_{2},g)=(f_{1}, g)+(f_{2}, g) \quad \forall f_{1}, f_{2}, g \in X$.
~\\
 
Specifically, then, a \emph{Hilbert space} $\mathcal{H}$ is an inner product space that is complete when furnished with the norm $||f||=\sqrt{(f,f)}$ \cite{Streater}. The first of the conditions for the scalar product is not a direct consequence of the principle of least radix economy and can be relaxed to include isotropic vectors (spinors). The Riesz-Fischer theorem \cite{Dym} establishes the isomorphism between the Hilbert space $\ell^{2}$ and the one of square summable real functions $L^{2}$. Between those spaces there exists a 1:1 linear distance-preserving map which constitutes the efficient implementation of the Fourier transform. All this also warrants that members in Eq. (\ref{ortonor}) constitute a complete orthonormal family in $\mathcal{H}$ 
\begin{eqnarray}
(e_{k'} ,e_{k} )&=& \int_{\eta}^{\eta+1}e^{-i2\pi (k'-k)\{S/h\}}d\left(S/h\right)= \delta_{kk'} \label{ortonorb}
\end{eqnarray}
that we can use as base of the vector space. In turn, this implies that the trigonometric series given by Eq. (\ref{genform}), $\psi(S)=\sum_{k=-\infty}^{\infty}\widetilde{\psi}(k)e^{i2\pi k S/h}=\sum_{k=-\infty}^{\infty}\widetilde{\psi}(k)e^{i2\pi k (\lfloor S/h \rfloor + \{S/h\})}=\sum_{k=-\infty}^{\infty}\widetilde{\psi}(k)e^{i2\pi k \{S/h\}}$ is also a Fourier series and that the latter is \emph{unique} for any set of elements $\widetilde{\psi}(k)$ in sequence space. We, therefore, have
\begin{equation}
\widetilde{\psi}(k)=\int_{\eta}^{\eta+1} \psi(S) e^{-i2\pi k\{S/h\}}d\left(S/h\right) = (e_{k}, \psi(S) ) \label{Fourier}
\end{equation}
and Eq. (\ref{genform}) can be rewritten as
\begin{equation}
\psi(S)=\sum_{k=-\infty}^{\infty}  (e_{k}, \psi(S) ) e_{k}
\end{equation}
which corresponds to a \emph{linear superposition} of the quantum states $e_{k}$. Because physical quantities that are observable can only be real-valued, since $\eta$ is real, as well as $S/h$ in the principle of least radix economy, finding a particular state $e_{k}$ of $\psi(S)$ is a just a possibility that may or may be not the case. We have $|(\psi(S), \psi(S) )|^{2}=1$ and, then, necessarily
\begin{equation}
|(e_{k}, \psi(S) )|^{2} \le 1 
\end{equation}
is a real quantity between $0$ and $1$ and since this is the only real-valued information that can be generally extracted out of $e_{k}$ and $\psi(S)$ it is reasonable to interpret $|(e_{k}, \psi(S) )|^{2}$ as the \emph{probability} that $e_{k}$ is the case in any measurement (Born rule). This is so because $\psi(S)$ represents the physical state of the system and therefore $|(e_{k}, \psi(S) )|^{2}$ is the norm of the projection of the physical state on the direction spanned by the vector $e_{k}$. 

%Since any value for the action is possible, we can also ask what is the probability of finding paths of Type I or Type II (which depends on whether $S/h$ is integer or not and hence, on the probability distribution of $\{ S/h \}$ as a function of $\eta$). In Section \ref{statistics} we address this question.

Taking into account the above remarks on probability we note then that $|(\psi(S), \psi(S) )|^{2}=1$ because a point in spacetime belongs always \emph{with certainty} to (at least) one path with action $S$ regardless of the actual value of $S$. This latter statement is clear both for periodic orbits (where any value of $S$ can be realized by completing the trajectory as many times as needed) and non-periodic ones (where any value of $S$ can be realized by extending the trajectory forward or backward in time). As discussed at the end of Section \ref{Schro}, the endpoints of the trajectory are irrelevant in the formulation presented here and we are only concerned with each point in spacetime $(\mathbf{q}, t)$ and the set of paths with action $S/h$ that pass through $(\mathbf{q}, t)$.

The consistency of all above results is still reinforced by the fact that the Dirichlet kernel $\mathcal{D}_{\infty}(2\pi S/h)$ is also everywhere a solution of the variational principle, Eq. (\ref{vari}). The Dirichlet kernel also obeys the symmetry Eq. (\ref{sime}) i.e. $\mathcal{D}_{\infty}(2\pi(S+nh)/h)=\mathcal{D}_{\infty}(2\pi S/h)$ and is to be regarded as an operator under an integral sign \cite{Lanczos} connecting in the circle $S^{1}$ the different values of the action $S/h$ which are consistent with the same wavefunction and for which the physical radix $\eta$ is kept constant. The Dirichlet kernel thus acts as a propagator between such states in the form of a convolution with the wavefunction 
\begin{equation}
    \psi(S)=\int_{\eta}^{\eta+1} \psi(S')\mathcal{D}_{\infty}\left(2\pi(S-S')/h\right)\,d(S'/h) \label{prop}
\end{equation}
It is well-known that the $L^{1}$ norm of the Dirichlet kernel $\mathcal{D}_{n}$ diverges as $\| D_n \| _{L^1} \sim \log n $ when $n \to \infty$. Therefore, in order to be able to use a function $\psi (S)$ in a convolution with the Dirichlet kernel, $\psi(S)$ must belong to the set of square-summable functions. This is, however, automatically warranted by the above development, which led us to establish the trigonometric series Eq. (\ref{genform}) as a Fourier series.

Quantum mechanics is thus described by unit rays $\psi$ (so-called because their norm $||\psi ||=\sqrt{(\psi, \psi)}=1$ \cite{Streater}) and the action of self-adjoint operators $\hat{O}$ which 
satisfy $(\psi_{B},\hat{O}\psi_{A})=\overline{(\hat{O}\psi_{A},\psi_{B})}$ so that the properties of the scalar product are unaffected and allow \emph{observables} (that are \emph{real-valued} quantities) to be defined. Since $\psi$ is a \emph{vector} in the space spanned by the complete orthonormal base Eq. (\ref{ortonor}), such operators are \emph{hermitian matrices} which send vectors to vectors and whose elements $[\hat{O}]_{ij}\equiv O_{ij}$ satisfy $O_{ij}=O_{ji}^{*}$. A most simple example is $\hat{K}$, the \emph{index operator}, defined as
\begin{equation}
\hat{K} \equiv \frac{\hbar}{i}\frac{d}{dS} \label{index1}
\end{equation}
We have, by using Eq. (\ref{genform})
\begin{equation}
\hat{K}\psi \equiv \frac{\hbar}{i}\frac{d\psi}{dS}=\sum_{k=-\infty}^{\infty}\widetilde{\psi}(k)\frac{\hbar}{i}\frac{d e^{i kS/\hbar}}{d S}=\sum_{k=-\infty}^{\infty}k\widetilde{\psi}(k)e^{i kS/\hbar} \label{index2}
\end{equation}
When this operator acts on wavefunctions which correspond to members of the orthonormal base, Eq. (\ref{ortonor}) we find
\begin{equation}
\hat{K}e_{k}=ke_{k} 
\end{equation}
which shows that any of such members is an eigenstate of the operator $\hat{K}$ with eigenvalue $k$. The matrix elements of the operator when the eigenfunctions are used as base are thus $(e_{k'}, \hat{K}e_{k})=k\delta_{kk'}$. This trivially shows that the operator $\hat{K}$ is hermitian.

In order to proceed further, we have now at our disposal the Hilbert space constructed through the \emph{complete} orthonormal base given by Eq. (\ref{ortonor}). Since all Hilbert spaces \emph{with the same dimensions} are \emph{isomorphic} \cite{Halmos} we are now free to choose any appropriate base in Hilbert space to deal with any particular quantum mechanical problem. This freedom in the choice of the base has been recently emphasized (see \cite{HooftNEW}, p. 176) and is fully consistent with the orthodox Copenhagen interpretation of quantum mechanics. For example, since $S$ is a scalar functional and can be thought as dependent on the generalized position vector $\mathbf{q}$ and time $t$, appropriate orthonormal functions other than the $e^{ikS/\hbar}$ can be chosen and this norm will always be conserved from Parseval's theorem.

We conclude this section with a brief summary of the main predictions made by the principle of least radix economy. There exist basically two kinds of physical solutions. Type I paths above lead to trajectories defined by the Euler-Lagrange \emph{differential} equations Eq. (\ref{Euler}), the state of the system being specified by the vector $(t, \mathbf{q})$. Type II paths are described by the \emph{integral} equation Eq. (\ref{prop}) in terms of the Dirichlet kernel, which acts as a propagator. \emph{In any case the physical state is described by the wavefunction $\psi(S)$ Eq. (\ref{genform}) with the discrete symmetry of Eq.(\ref{sime})}. In the classical limit $S/h$ large only type I paths are relevant. In the quantum regime, the fact that the two types of paths coexist in phase space is consistent with the wave-particle duality. In the case $S/h$ integer, each term in the sum in Eq. (\ref{genform}) describes a de Broglie standing wave in the quantum regime.

\section{Derivation of the Schr\"odinger equation, eikonal approximation and correspondence principle} \label{Schro}

From Eq. (\ref{genform}) it is now straightforward to derive the Schr\"odinger equation. We first define $S_{k} \equiv k S$, $\mathbf{p}_{k} \equiv k\mathbf{p}$ and $E_{k} \equiv k H$. Then Eq. (\ref{HJ}) implies 
\begin{equation}
\ E_{k}
=- {\partial S_{k} \over \partial t} \qquad \ \ \mathbf{p}_{k}=\nabla S_{k} \label{HJ2}
\end{equation}
and, therefore, from Eq. (\ref{genform})
\begin{eqnarray}
\frac{\partial \psi}{\partial t}&=&\sum_{k=-\infty}^{\infty}\widetilde{\psi}(k)\frac{\partial e^{ikS/\hbar}}{\partial t}=\frac{i}{\hbar}\sum_{k=-\infty}^{\infty}\widetilde{\psi}(k)e^{iS_{k}/\hbar}\frac{\partial S_{k}}{\partial t}  \nonumber \\
&=& -\frac{i}{\hbar} \sum_{k=-\infty}^{\infty}E_{k}\widetilde{\psi}(k)e^{iS_{k}/\hbar} \equiv -\frac{i}{\hbar} \widehat{H} \psi
\label{genform2}
\end{eqnarray}
\begin{eqnarray}
\nabla \psi &=&\sum_{k=-\infty}^{\infty}\widetilde{\psi}(k)\nabla e^{ikS/\hbar} 
=\frac{i}{\hbar}\sum_{k=-\infty}^{\infty}\widetilde{\psi}(k)e^{iS_{k}/\hbar}\nabla S_{k}  \nonumber \\
&=& \frac{i}{\hbar} \sum_{k=-\infty}^{\infty}\mathbf{p}_{k}\widetilde{\psi}(k)e^{iS_{k}/\hbar} \equiv \frac{i}{\hbar} \widehat{\mathbf{p}} \psi
\label{genform3} \\
\nabla^{2} \psi &=& \nabla \cdot \nabla \psi= -\frac{1}{\hbar^{2}} \sum_{k=-\infty}^{\infty}\mathbf{p}_{k}^{2}\widetilde{\psi}(k)e^{iS_{k}/\hbar} \nonumber \\ 
&=& -\frac{1}{\hbar^{2}} \widehat{\mathbf{p}}\cdot \widehat{\mathbf{p}} \psi
\label{genform4}
\end{eqnarray} 
In Eqs. (\ref{genform2}) and (\ref{genform3}) the Hamiltonian $\widehat{H} \equiv i\hbar\frac{\partial}{\partial t}$ and momentum operators $\widehat{\mathbf{p}} \equiv -i\hbar\nabla$ have been defined. From Eqs. (\ref{genform2}) and (\ref{genform4}) we now have
\begin{eqnarray}
&& i\hbar\frac{\partial \psi}{\partial t}+\frac{\hbar^{2}}{2m}\nabla^{2} \psi =
\sum_{k=-\infty}^{\infty}\left(E_{k}-\frac{\mathbf{p}_{k}^{2}}{2m}\right)\widetilde{\psi}(k)e^{iS_{k}/\hbar} \nonumber \\
&&= \sum_{k=-\infty}^{\infty}V_{k}(\mathbf{q})\widetilde{\psi}(k)e^{iS_{k}/\hbar}=V(\mathbf{q})\psi
\end{eqnarray}
where $V(\mathbf{q})$ is the potential energy and it has been used that $E_{k}=\frac{\mathbf{p}_{k}^{2}}{2m}+V_{k}(\mathbf{q})$ (conservation of energy). We thus obtain the time-dependent Schr\"odinger equation
\begin{equation}
i\hbar\frac{\partial \psi}{\partial t}=\left(-\frac{\hbar^{2}}{2m}\nabla^{2}+V(\mathbf{q})\right)\psi=\widehat{H} \psi \label{Sch}
\end{equation}
If we consider a free particle ($V(\mathbf{q})=0$) we obtain from Eqs. (\ref{genform}) and (\ref{Sch}) the following solution for $\psi(S)$
\begin{equation}
\psi(S)=\sum_{k=-\infty}^{\infty}\widetilde{\psi}(k)e^{iS_{k}/\hbar}=Ae^{i (\mathbf{p}\mathbf{q}-Et)/\hbar}
\end{equation}
with $A$ being a constant. This plane wave corresponding to the free particle can be interpreted from this latter expression as a mean-field (averaged) complex order parameter of an (infinite) collection of ``oscillators''  $\widetilde{\psi}(k)e^{iS_{k}/\hbar}$. Each point in spacetime can thus be assumed to contain such an infinite collection of oscillators which are \emph{not} to be considered as hidden variables: Only their mean field is physically relevant and, furthermore, only the power spectral density of the order parameter is physically observable.

Since Eq. (\ref{genform}) converges everywhere on a set of positive Lebesgue measure, by the Cantor-Lebesgue theorem $\widetilde{\psi}(k) \to 0$ as $|k|\to \infty$ \cite{Bruckner}. We use now this latter result to prove that \emph{when $S/h$ is increasingly large, in the semiclassical regime Eq. (\ref{genform}) becomes}
\begin{equation}
\psi(S) \sim \widetilde{\psi}(1) e^{iS/\hbar}  \label{eikonal}
\end{equation}
To prove this we use Eq. (\ref{eqmode1}) in Eq. (\ref{genform}) 
\begin{equation}
\psi(S)=\sum_{k=-\infty}^{\infty}\widetilde{\psi}(k)e^{i2\pi kS/h}= \sum_{k=-\infty}^{\infty}\widetilde{\psi}(k)e^{i2\pi k(\eta+1)\left \{\frac{S/h}{\eta+1}\right \}}
\label{genformIIb}
\end{equation}
Thus $S/h$ large corresponds to taking the limit $\eta+1$ large. Now, note that from Eq. (\ref{Fourier})
\begin{equation}
\widetilde{\psi}(k(\eta+1))=\int_{0}^{1} \psi\left(\frac{S/h}{\eta+1} \right) e^{-i2\pi k(\eta+1)\left \{\frac{S/h}{\eta+1}\right \}}d\left(\frac{S/h}{\eta+1}\right) \label{Fourier2}
\end{equation}
Let $k' \equiv k(\eta+1)$. Then, because of the Cantor-Lebesgue theorem, $\widetilde{\psi}(k') \to 0$ as $k' \to \infty$, and for $\eta \to \infty$ the only $k$'s that are relevant are the ones that become smaller accordingly, i.e. long wavelengths. Thus, asymptotically only the first inhomogeneous mode ($|k|=1$) present in the expansion Eq. (\ref{genformIIb}) becomes significant, which in turn proves Eq. (\ref{eikonal}). If one replaces Eq. (\ref{eikonal}) in the Schr\"odinger equation one arrives at the Hamilton-Jacobi equation in the limit $S/\hbar$ large. Mathematically, we can see this explicitly if we replace $\psi=e_{1}(S/h)=e^{iS/\hbar}$ in the Schr\"odinger equation. We then obtain
\begin{equation}
\frac{(\nabla S)^{2}}{2m}+V(\mathbf{q})+\frac{\partial S}{\partial t}=\frac{i\hbar}{2m}\nabla^{2}S \label{HJESch}
\end{equation}
The r.h.s. of Eq. (\ref{HJESch}) vanishes because the prefactor (of order $\hbar$) is negligible compared to the prefactors of all terms in the l.h.s (of order unity). We are thus only left with
\begin{equation}
\frac{(\nabla S)^{2}}{2m}+V(\mathbf{q})+\frac{\partial S}{\partial t}=0 \label{HJE2}
\end{equation}
which is the Hamilton-Jacobi equation, Eq. (\ref{HJE}), for a Hamiltonian of the form $H=\frac{\mathbf{p}^{2}}{2m}+V(\mathbf{q})$. This proves again the correspondence principle: \emph{in the limit of large quantum numbers, quantum mechanics reproduces classical mechanics}. 

Note that time reversal symmetry is possessed by the Schr\"odinger equation, i.e., we have, from Eq. (\ref{Sch}), by taking the complex conjugate and using that $H$ is real
\begin{equation}
\widehat{H} \psi^{*}=
-i\hbar\frac{\partial \psi^{*}}{\partial t}=i\hbar\frac{\partial \psi^{*}}{\partial (-t)}    \label{Sch2}
\end{equation} 
Thus, the state $\psi^{*}$ will evolve in the $+t$ direction exactly in the same way as $\psi$ would have evolved in the $-t$ direction. This warrants that the probability density, proportional to $|\psi|^{2}$, remains unaffected \cite{Tinkham}.

Although the formulation presented here is based on the Lagrangian action $S(\mathbf{q}(t))$ our approach is different to Feynman's path integral formulation of quantum mechanics \cite{Feynman}. Note that our formulation is entirely built on the wavefunction Eq. (\ref{genform}), which is itself a consequence of the least radix economy principle: there is not a ``sum over all possible different paths in spacetime''. Our formulation is based neither on the concept of propagator between points in spacetime nor on attributing amplitude probabilities to paths (but to physical states directly). In Feynman's formulation the endpoints $\mathbf{q'}$ and $\mathbf{q''}$ of the trajectory are fixed and any midpoint $\mathbf{q}(t)$ and the action $S(\mathbf{q}(t))$ fluctuate. A sum over all paths is then needed to obtain the probability amplitude of going from $\mathbf{q'}$ to $\mathbf{q''}$. In our formulation, we rather fix $S$ and $\mathbf{q}(t)$ and we do not care about the endpoints, which are arbitrary in the quantum regime. Indeed, there is in general an ensemble of endpoints (and, hence, an ensemble of paths, see Fig. \ref{paths}) that are compatible with a value for the action $S(\mathbf{q}(t))$ at $\mathbf{q}(t)$. Importantly, note however that, in the classical regime, the trajectory is uniquely specified by giving $\mathbf{q}(t)$ and $S(\mathbf{q}(t))$ (since then the momentum is uniquely determined along the trajectory through the Hamilton-Jacobi equation and the total length of the path is given by $S$). We have then shown how the Hilbert space emerges as a natural description when $\eta$ is small.

\begin{figure}
\includegraphics[width=0.3 \textwidth, angle=270]{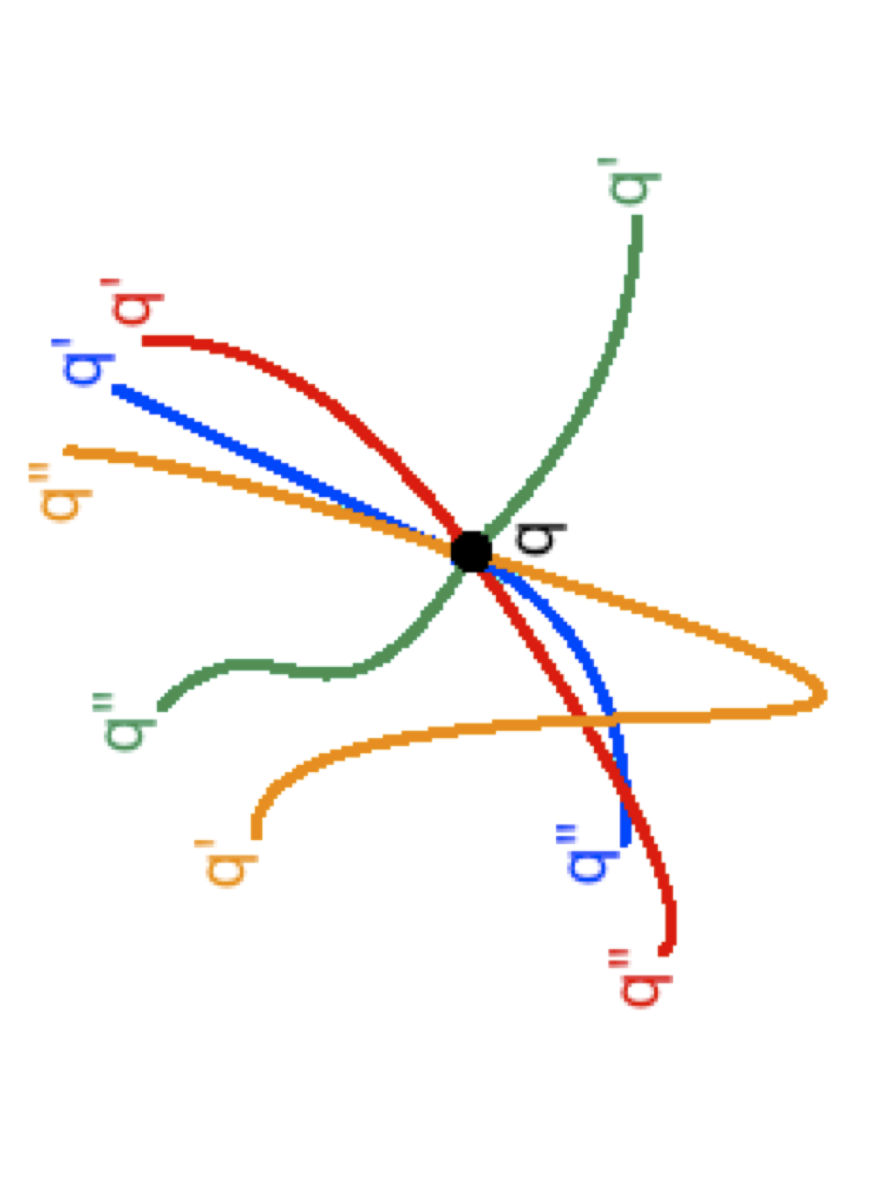}
\begin{center}
\caption{The action $S(\mathbf{q}(t))$ at a point $\mathbf{q}$ is a real number that may refer to any of the many different paths in spacetime (not necessarily least-action ones) with (unspecified) initial $\mathbf{q'}$ and final $\mathbf{q''}$ coordinates (the paths being parameterized by the time coordinate $t$).} \label{paths}
\end{center}
\end{figure}

We have made use of Eqs. (\ref{HJ2}) and this requires some justification since the Schr\"odinger equation is thus to be understood as an approximation related to an expansion around the least-action paths. To see this let us observe that, in general, we have \cite{Gutzwiller}
\begin{equation}
S(\mathbf{q}(t))\equiv S(\mathbf{q'} \to \mathbf{q} \to \mathbf{q''})=S(\mathbf{q'} \to \mathbf{q})+S(\mathbf{q} \to \mathbf{q''}) \label{trans1}
\end{equation}
where we have split the total action into partial actions and the transitions between the endpoints through the midpoint $\mathbf{q}$ are indicated. Let Eq. (\ref{trans1}) refer to a least-action path. Then, for an arbitrary path (\emph{not} a least-action one) between the same points, we would have
\begin{eqnarray}
&&S(\mathbf{q}(t)+\delta \mathbf{q}) \equiv S(\mathbf{q'} \to \mathbf{q}+\delta \mathbf{q} \to \mathbf{q''})=S(\mathbf{q'} \to \mathbf{q}+\delta \mathbf{q})+S(\mathbf{q}+\delta \mathbf{q} \to \mathbf{q''}) \nonumber \\
&&= S(\mathbf{q'} \to \mathbf{q})+S(\mathbf{q} \to \mathbf{q''})+\frac{1}{2}\delta q_{j}\left(\frac{\partial^{2} S(\mathbf{q'} \to \mathbf{q})}{\partial q_{j}\partial q_{k}}+\frac{\partial^{2} S(\mathbf{q} \to \mathbf{q''})}{\partial q_{j}\partial q_{k}} \right)\delta q_{k} \nonumber \\
&&=S(\mathbf{q}(t))+\frac{1}{2}\delta q_{j}c_{jk}\delta q_{k}
\end{eqnarray}
where we have used Einstein's summation convention and that the first variation around the classical path is zero. We have also defined 
\begin{equation}
c_{jk} \equiv \frac{\partial^{2} S(\mathbf{q'} \to \mathbf{q})}{\partial q_{j}\partial q_{k}}+\frac{\partial^{2} S(\mathbf{q} \to \mathbf{q''})}{\partial q_{j}\partial q_{k}}
\end{equation}
These latter coefficients form a matrix which gives the deviation from the extremum of the action. If the second variation is small so are these coefficients and the determinant of the matrix is small as well. Indeed, the density of trajectories close to the classical one is inversely proportional to the  determinant of this matrix (see \cite{Gutzwiller}, p. 179): a larger density is related to a constructive interference of waves \cite{Gutzwiller} and there is where the approximation involved in keeping to first-order to derive the Schr\"odinger equation (hence using Eqs. (\ref{HJ2}) to first-order) is most accurate. This approximation is also involved in Feynman's derivation of the Schr\"odinger equation \cite{Feynman} (see also the remarks in \cite{Gutzwiller}) and this shows the intimate connection between Feynman's formulation and ours, even when we are not following a path integral approach. Rather, we claim that Eq. (\ref{genform}) captures the whole interference pattern caused by the paths of total action $S$ passing through the location $\mathbf{q}(t)$. The intensity of this interference of waves at $\mathbf{q}(t)$ is thus naturally related to the probability of finding $\mathbf{q}(t)$ as a physical state of the system.

\section{Breaking of spacetime commutativity and finite systems} \label{noncom}

If we define the position operator as $\widehat{\mathbf{q}} \equiv \mathbf{q}$, we have, for two components $i$ and $j$ $[\widehat{q}_{i},\widehat{p}_{j}]\psi=\left(\widehat{q}_{i}\widehat{p}_{j}-\widehat{p}_{j}\widehat{q}_{i}\right)\psi=-i\hbar\left(q_{i}\frac{\partial}{\partial q_{j}}-\frac{\partial}{\partial q_{j}}q_{i}\right)\psi=i \hbar \delta_{ij}\psi$ (with $\delta_{ij}$ being the Kronecker delta).
In general, for two conjugate variables $\alpha$ and $\beta$ that satisfy $\beta_{k}=\partial S_{k}/\partial \alpha$, the operator $\widehat{\beta}=-i\hbar \partial/\partial \alpha$ can be defined through the straightforward generalization of Eq. (\ref{genform3}). Then, if one considers the operator $\widehat{\alpha} \equiv \alpha$, the commutation relationship $[\widehat{\alpha},\widehat{\beta}]=i\hbar$ holds (provided that $\alpha$ is differentiable and periodic in the unit circle $S^{1}$). This implies the Heisenberg uncertainty principle \cite{Kennard} \cite{Robertson}. The breaking of geometric commutativity has been the subject of intense interest \cite{Kauffman1} \cite{Connes} and can also be directly understood from the concepts introduced in this article. If we now use $\eta$ given by Eq. (\ref{etadef}) as radix in Eq. (\ref{fund1}), we have, for $A=S/h$ 
\begin{equation}
\frac{S}{h}=1\cdot \eta^{1}+0\cdot \eta^{0}+ \sum_{m=-\infty}^{0} \eta^{m-1}\mathbf{d}_{\eta}(m, S/h) \label{fundS}
\end{equation} 
For classical paths for which the action is large the sum in the last term can be neglected: Eq. (\ref{fundS}) means the same as $S/h=\eta+\{S/h\}$, and $\{S/h\}$ can be neglected for $\eta$ large. In this classical limit we are left with only two digits at integer positions: All digits after the decimal point are zero. From Eq. (\ref{digit}) we have, for each of these two digits, $\mathbf{d}_{\eta}(1, S/h)=0= \left \lfloor S/h \right \rfloor - \eta \left \lfloor \frac{S/h}{\eta}\right \rfloor$, $\mathbf{d}_{\eta}(2, S/h)=1= \left \lfloor \frac{S/h}{\eta} \right \rfloor - \eta \left \lfloor \frac{S/h}{\eta^{2}}\right \rfloor = \left \lfloor \frac{S/h}{\eta} \right \rfloor$. The former of these equations means that the operations of dividing by $\eta$ and taking the floor brackets $\left \lfloor ... \right \rfloor$ commute, and hence $\left \lfloor (S/h)/\eta\right \rfloor=\left \lfloor S/h \right \rfloor /\eta$. The latter equation means that $S/h$ is proportional to $\eta$ with prefactor 1. Let us now assume, however, than we are in the quantum regime so that Eq. (\ref{crit}) is satisfied. This means that $S/h \sim \{ S/h \}$ and thus the fractional part cannot be neglected (the sum in Eq. (\ref{fundS}) contains non-zero terms). Let us assume that in the fractional part of $S/h$ the digit $\mathbf{d}_{\eta}(-|m'|, S/h)$ accompanying the power $\eta^{-|m'|-1}$ with $m' \le 0$ in Eq. (\ref{fundS}) is the first, most significant, nonzero digit. Then from Eq. (\ref{digit}) we have
\begin{equation}
\mathbf{d}_{\eta}(-|m'|, S/h) = \left \lfloor \eta^{1+|m'|} S/h \right \rfloor - \eta \left \lfloor \eta^{|m'|}S/h \right \rfloor \label{digitfin}
\end{equation}
This latter expression means that the operator $\hat{\eta}x \equiv \eta x$ that multiplies the real quantity $x$ by $\eta$ and the operator $\lfloor \ldots \rfloor x \equiv \lfloor x \rfloor$ which evaluates the floor function \emph{do not} commute (in Eq. (\ref{digitfin}) both operators act on the quantity $\eta^{|m'|}S/h$). The classical limit (where these operators do commute) occurs in the limit $|m'| \to \infty$ (i.e. when the fractional part $\{ S/h \}$ is negligible). In such limit, the principle of least radix economy reduces to the principle of least action and commutativity is regained. Note that this non-commutative relationship for the action is not a conventional Heisenberg-like as the ones discussed above since $\mathbf{d}_{\eta}(-|m'|, S/h)$ in Eq. (\ref{digitfin}) is an integer number $\in [0, \eta-1]$. Such integer-valued structure constants arise in the study of Chevalley groups \cite{Chevalley}. The latter are non-abelian finite simple groups that constitute the finite counterparts of Lie groups \cite{Carter}. 

In atomic models, the physical radix $\eta$ coincides with the so-called \emph{principal quantum number} $n$. Then it must also be remarked that $d_{\eta}(m, S/h)$ for \emph{any} $m \le 0$ is a non-negative integer $\in [0, \eta-1]$. This suggests that $d_{\eta}(-|m'|, S/h)$ (i.e. the most significant digit of the fractional part of the action $\{ S/ h \}$) corresponds to the \emph{azimuthal quantum number} $\ell$ describing the orbitals (electronic subshells). The necessary \emph{existence} of such quantum numbers comes directly from the variational principle presented in this article without solving any further equation. When the angular momentum is important as a further conservation law coming from the semiclassical problem, the splitting of the main quantum shells into orbital subshells predicted by the Schr\"odinger equation, Eq. (\ref{Sch}), can also be understood from this new point of view as the increased significance that the digit $d_{\eta}(-|m'|, S/h)$ acquires on the expansion of the Lagrangian action $S/h$ in the radix $\eta$, Eq. (\ref{fundS}). \emph{That this digit is responsible for the breaking of the commutativity of the spacetime, from Eq. (\ref{digitfin}) is also made spatially evident, since orbitals are countable discrete objects that arise out of a continuum and commutative spacetime.} 

This argument points to a natural way in which finite systems with a finite number of quantum states enter in the theory. Let us first note that, from Eq. (\ref{fundS}) we have
\begin{eqnarray}
\psi(S)&=&\sum_{k=-\infty}^{\infty}\widetilde{\psi}(k)e^{ikS/\hbar}=\sum_{k=-\infty}^{\infty}\widetilde{\psi}(k)e^{i2\pi k(\lfloor S/h \rfloor +\{ S/h \})}  \label{gen2} \\
%&=& \sum_{k=-\infty}^{\infty}\widetilde{\psi}(k)e^{i2\pi k\{ S/\hbar \}} \nonumber \\
&=& \sum_{k=-\infty}^{\infty}\widetilde{\psi}(k)\exp \left(i2\pi k\sum_{m=-\infty}^{0} \eta^{m-1}\mathbf{d}_{\eta}(m, S/h)\right) \nonumber
\end{eqnarray}

Let us now assume a physical system with a phase space such that \emph{all digits in the expansion of the fractional part of the action in Eq. (\ref{fundS}) can be neglected except} $d_{\eta}(0, S/h) \in [0, \eta -1]$. In this case, we obtain
\begin{eqnarray}
\psi(S)&=& \sum_{k=-\infty}^{\infty}\widetilde{\psi}(k)\exp \left(i\frac{2\pi k \mathbf{d}_{\eta}(0, S/h)}{\eta}\right) \nonumber \\
&=&\sum_{k'=0}^{\eta-1}\varphi (k')\exp \left(\frac{i2\pi k'\mathbf{d}_{\eta}(0, S/h) }{\eta}\right)\equiv \psi(\mathbf{d}_{\eta}(0, S/h) )
\end{eqnarray}
where we have defined $\varphi (k')\equiv \sum_{k=-\infty}^{\infty}\widetilde{\psi}(k'+k\eta)$. Thus $\psi(\mathbf{d}_{\eta}(0, S/h))$ is a wavefunction in the finite-dimensional Hilbert space $\mathbb{C}^{\eta}$. There is a vast literature addressing wavefunctions which constitute finite models of this kind, see e.g. \cite{HooftNEW} and \cite{Schwinger}. Following \cite{HooftNEW} we can, for example, construct a Hamiltonian operator $\widehat{H}$ which acts on the wavefunction $\psi(\mathbf{d}_{\eta}(0, S/h) )$ as
\begin{equation}
 \widehat{H}\psi(\mathbf{d}_{\eta}(0, S/h) )=\frac{2\pi \hbar \mathbf{d}_{\eta}(0, S/h) }{\eta \epsilon}\psi(\mathbf{d}_{\eta}(0, S/h) ) \label{spectrum}
\end{equation}
where $\epsilon$ is a fundamental time step. The time interval $\Delta t$ is regarded here as an integer multiple of that fundamental time step. Such a model with a discrete clock belongs to the family of cellular automata models \cite{VGM1}, \cite{HooftNEW}. This is equivalent to the cogwheel model with $\eta$ teeth described in Section 2.3 in \cite{HooftNEW}. From the Schr\"odinger equation, we then have
\begin{eqnarray}
\psi(\mathbf{d}_{\eta}(0, S/h) , t=\epsilon)&=&e^{-i\widehat{H}\epsilon/\hbar}\psi(\mathbf{d}_{\eta}(0, S/h) ,t=0) \nonumber \\
&=&\sum_{k'=0}^{\eta-1}\varphi (k')\exp \left(\frac{i2\pi (k'-1)\mathbf{d}_{\eta}(0, S/h) }{\eta}\right)  \\
&=&\sum_{k'=0}^{\eta-1}\varphi (k'+1 \mod \eta)\exp \left(\frac{i2\pi k' \mathbf{d}_{\eta}(0, S/h) }{\eta}\right) \nonumber
\end{eqnarray}
The spectrum given by Eq. (\ref{spectrum}) is ubiquitously found in physics \cite{HooftNEW} and corresponds to the one of an atom with total angular momentum $J=\frac{1}{2}(\eta-1)$ and magnetic moment $\mu$ in a weak magnetic field: the Zeeman atom \cite{HooftNEW}. From the above we also see that $\psi(\mathbf{d}_{\eta}(0, S/h) )$ can be written as a linear superposition of the $\eta$ members $e_{k'}(\mathbf{d}_{\eta}(0, S/h) /\eta)$ $(k' \in [0, \eta-1]$, which correspond to rational values $\mathbf{d}_{\eta}(0, S/h) /\eta$ of the action.

\section{Statistics of action} \label{statistics}

In any given physical situation the actual radix can \emph{fluctuate} around the optimal one (having the least economy) because all values for the action are allowed by the principle of least radix economy. In the classical limit (least action paths) the actual radix and the optimal one can be taken as equal because the impact of fluctuations in the actual radix can be neglected. Let us now investigate these statements more closely. 

When the optimal radix is $\eta=\eta_{1}=1$ (unary radix) we can ask what is the probability to observe either a quantum of action (a '1' value) or a vacuum state (a '0' value) which has \emph{always} (trivially) the least economy. In the deep quantum regime, \emph{since no other information is available from the principle}, the statistics of action reduces to the analysis of unbiased yes/no experiments. Such a kind of statistics is well-known to be modeled by a binomial distribution of the form
\begin{equation}
f(m; n,1/2)={n \choose m}\frac{1}{2^{n}} \label{probgene}
\end{equation}
The latter gives the probability that out of $n$ experiments $m$ quanta of action (and, hence, $n-m$ vacuum states) are observed. 

We can now ask how does this distribution change when the optimal radix $\eta$ is larger than one. In order to investigate this question, we first note that, if in the binomial distribution the probability of observing a quantum of action in one experiment is now $p$, we have
\begin{equation}
f(m; n,p)={n \choose m}p^{m}(1-p)^{n-m} \label{binomial}
\end{equation}
for the probability of observing $m$ quanta in $n$ experiments. The mode (i.e. most probable value) of Eq. (\ref{binomial}) is
\begin{equation}
\lfloor (n+1) p \rfloor
\end{equation} 
when $(n+1)p$ is noninteger. We then observe that if we identify $n$ with $\eta$ and $p$ with $\left\{ \frac{S}{(\eta+1)h} \right \}$ then, from Eqs. (\ref{eqmode}) and Eq. (\ref{binomial}) we find that $\eta$ is indeed the mode of the binomial distribution given by
\begin{equation}
f\left(m; \eta, \left\{ \frac{S}{(\eta+1)h} \right \} \right)={\eta \choose m}\left\{ \frac{S}{(\eta+1)h} \right \}^{m}\left(1-\left\{ \frac{S}{(\eta+1)h} \right \}\right)^{\eta-m} \label{binomech}
\end{equation}
i.e., this probability mass function has its maximum at value $m=\eta$ which appears with probability
\begin{equation}
f\left(\eta; \eta, \left\{ \frac{S}{(\eta+1)h} \right \} \right)=\left\{ \frac{S}{(\eta+1)h} \right \}^{\eta} \label{maxim}
\end{equation}
The mean and the variance of the distribution are given respectively by  $\eta\left\{ \frac{S}{(\eta+1)h} \right \}$ and
$\eta\left\{ \frac{S}{(\eta+1)h} \right \}\left(1-\left\{ \frac{S}{(\eta+1)h} \right \} \right)$. The relative strength of the fluctuations around the maximum at $\eta$ is therefore given by
\begin{equation}
\sqrt{\frac{1-\left\{ \frac{S}{(\eta+1)h} \right \} }{\eta\left\{ \frac{S}{(\eta+1)h} \right \}}} \sim \frac{a}{\sqrt{\eta}} \label{fluctumech}
\end{equation}
where $a=O(1)$ because of Eq. (\ref{desigul}). Thus, the distribution of action is strongly peaked at the optimal radix $\eta$ in the limit $\eta$ large. Although the outcomes $\eta-1$, $\eta-2$ etc. are also possible with a significant probability, in such limit the system becomes increasingly better described by $\eta$ on the average because any other possible integer values can be accurately approximated by $\eta$. \emph{That such approximation at the level of integers can be made in the classical limit, automatically warrants the validity of the variational approach of classical mechanics in that limit as well}. This provides an interpretation of how the principle of least-action emerges from the quantum world out of probabilistic considerations. We then see that the physical radix $\eta$ not only classifies physical paths giving the characteristic, optimal and most economic ruler of the dynamics: It also classifies the dynamical behavior into binomial probability distributions, Eq. (\ref{binomech}), of which it constitutes the mode. 

If $\eta$ is large there is an almost unit probability of observing a chain of $\eta$ 1's on the average. Thus the optimal radix $\eta$ corresponds to observing chains whose most probable outcome in $\eta$ 'experiments' are $\eta$ quanta of action. From Eq. (\ref{probgene}) the probability of observing $n$ quanta of action in the unary radix $\eta_{1}=1$ is $(1/2)^{n}$. 

In order to observe $n$ quanta of action in the unary radix \emph{with the most significant probability}, at least $2n$ experiments have to be carried out. Note that there are ${2n \choose n}=\frac{(2n)!}{(n !)^{2}}$ chains with $n$ zeroes and $n$ ones. For $2n$ experiments, these chains are the largest in number and, hence, they have the most significant probability, which is given by
\begin{equation}
\frac{1}{2^{2n}}{2n \choose n} \label{proba}
\end{equation} 
This means that, in a situation where the optimal radix is $\eta_{1}$, some chains of ten experiments yielding e.g.
\begin{equation}
0100111001 \qquad 1010011001 \qquad 1100101010  
\end{equation}
are all consistent with finding the chain
\begin{equation}
11111  
\end{equation}
in radix $\eta=5$ after performing five experiments. Note however that, while fluctuations around this maximum are stronger in $\eta_{1}$ (and, hence, chains with different numbers of ones and zeros have also a significant probability) they are less significant in radix $\eta=5$ from the above arguments. However, \emph{because of the law of large numbers in the classical limit ($\eta \to \infty$) chains of $\eta$ ones in radix $\eta$ are described in radix $\eta_{1}$ by chains with length $2\eta$ containing $\eta$ zeroes and $\eta$ ones: Any other kind of chains have negligible measure.} This statement, whose proof is the law of large numbers \cite{Kac}, implies then that Eqs. (\ref{maxim}) and (\ref{proba}) must coincide on the average and in the limit $\eta$ large fluctuations must be negligible in both the $\eta$ and the $\eta_{1}$ ``ensembles'', i.e. we have
\begin{equation}
p(\eta) \equiv \left< f\left(\eta; \eta, \left\{ \frac{S}{(\eta+1)h} \right \} \right)\right>=\frac{1}{2^{2\eta}}{2\eta \choose \eta}
\end{equation}
where the brackets denote the average over all possible values of the action. From this latter expression, we obtain the following result
\begin{equation}
2^{2\eta}p(\eta)=(\eta+1)C_{\eta} \label{supercatal}
\end{equation}
where $C_{\eta}$ is the Catalan number $n=\eta$. In the next section we shall discover what this Catalan number counts.

%=\eta\left\{ \frac{S}{(\eta+1)h} \right \}
%=\left<\left\{ \frac{S}{(\eta+1)h} \right \}^{\eta}\right>=
%\emph{Determinism ($\eta$ large) and purely random statistical behavior ($\eta \sim 1$ small) coexist in spacetime}. The fact that Eqs. (\ref{binomech}), (\ref{maxim}) and (\ref{fluctumech}) are \emph{not} invariant under the discrete symmetry Eq. (\ref{sime}) is crucially important. In fact, this result also helps to interpret the collapse of a quantum state in a measurement process. When a measurement takes place, the optimal radix $\eta$ of the quantum system suddenly changes to the optimal radix $\eta' >> \eta$ of the composite system formed by the quantum system and the (classical or semiclassical) measurement device. As a consequence, there is a change in the binomial distribution associated from the optimal radix $\eta$ to the larger one corresponding to $\eta'$. The latter is strongly peaked at  $S/h=\eta'$ and any quantum fluctuation around this value becomes negligible. 

The above discussion shows that $\eta$ plays an analogous role to the number of particles $N$ in classical statistical thermodynamics where fluctuations around mean values of thermodynamic quantities are proportional to $1/\sqrt{N}$. We emphasize that this result indeed \emph{is not} an analogy, as the next section shows. (There, the physical radix $\eta$ is indeed shown to be equivalent to $\eta$ \emph{particles} described by the unary radix.) With this in mind, the radix economy for the quantity $\mathcal{A}$ that enters the least radix economy principle is asymptotically equal to 
\begin{equation}
\mathcal{C}\left(\eta, \mathcal{A} \right)= \eta \left \lfloor1+\log_{\eta} \mathcal{A} \right \rfloor  \sim  
\eta \log_{\eta} \mathcal{A}= \log_{\eta} \mathcal{A}^{\eta}
\label{capaceta}
\end{equation}
when $\mathcal{A}$ is large. Eq. (\ref{capaceta}) has the form of a ``Boltzmann entropy'' or a ``Massieu-Planck thermodynamic potential'' \cite{VGMStat} and tells us that $\mathcal{A}$ is an ``extensive'' quantity in the ``number of particles'' $\eta$ when $\mathcal{A}$ is large. Thus, the radix capacity is, generally, an entropy-like quantity and the principle of least-radix economy leads to a relationship between action (for $\mathcal{A}=S/h$) and entropy. Such a relationship was explored, by totally different arguments, in \cite{Annals}. In Section \ref{entropy} we shall elucidate the connection between this sketch of ``spacetime thermodynamics'' and classical statistical thermodynamics. 

As a final remark of this section, we note that from Eq. (\ref{raduno}) we have
\begin{equation}
\mathcal{C}(1,\eta)=\eta
\end{equation}
Because of the above interpretation of the radix economy $\mathcal{C}$ as an entropy-like quantity, this relationship relates the action $\eta$ to the ``entropy $\mathcal{C}(1, \eta)$ of a single particle'' (described by the quantum of action in the unary radix) and coincides with the connection between action and entropy made by de Broglie in his book ``Thermodynamics of the isolated particle'' \cite{deBroglie1} and also discussed in \cite{deBroglie2}. Note that this relationship has a different form to the familiar one given by Eq. (\ref{capaceta}) but this is merely because the unary radix behaves quite differently to any other radix $\eta \ge 2$, as we explained in Section \ref{radie}.

\section{The unary radix: special relativity, relativistic wave equations and spin from the quantum of action.} \label{unary}

We now explore the consequences of Postulate \textbf{C} in Section \ref{PLRE}. As explained in Section \ref{radie}, the unary radix $\eta_{1}=1$ defines the most elementary numeral system and  it constitutes the lowest bound for a physical, nonvanishing radix. 

Postulate \textbf{C} implies that \emph{action is quantized} and that the quantum of action is described by the unary physical radix $\eta_{1}=\lfloor S/h \rfloor=1$. Any other value of $\eta$ can be itself written in the unary radix as
\begin{eqnarray}
 \begin{matrix} \eta = &  \underbrace{1_{}11\dots 1 1} \\ & \mbox{$\eta$ copies of 1} \end{matrix}
 \label{secondquan}
\end{eqnarray}
like in the examples given in Eq. (\ref{examples}) of Section \ref{radie}. Thus a situation described by optimal radix $\eta$, no matter how large, can be seen as involving $\eta$ quanta of action described each by radix $\eta_{1}$. This is consistent with the observation made about $\eta$ in the previous section that $\eta$ is analogous to a ``number of particles''. Indeed, it is the number of quanta of action.

The solution of the Schr\"odinger equation leads, generally, to an spectrum of discrete energy levels (e.g. for a particle in a potential well). When these are infinitesimally separated, a continuum energy spectrum is obtained instead. Still, we can think about energy in similar terms as we think about action: \emph{Energy is quantized as well} and we can  talk about \emph{energy quanta} in terms of the energy differences of consecutive energy levels. The ratio of a quantum of action to a quantum of energy is \emph{time}. Since the latter can be thought as measured by a clock that has an absolute character, we generally define a \emph{proper time interval} as the ratio of the quantum of action $\eta_{1}$ to the quantum of energy $E_{1}$    
\begin{equation}
\Delta T \equiv \frac{h\eta_{1}}{E_{1}}
\end{equation}
From the results in the previous section and the principle of least radix economy, we have seen that action can fluctuate. If we consider processes in which \emph{a quantum of energy is conserved while considering a transformation \emph{from} a quantum of action \emph{to} a different physical action}, then the ratio $\eta_{1}/\Delta T$ must stay constant, being equal to the ratio $S/(h\Delta t)$.  Here $\Delta t$ is a time increment characteristic of the process with the action $S/h \ne \eta_{1}$ (an improper time). Thus, we have 
\begin{equation}
\frac{\eta_{1}}{\Delta T}=\frac{S/h}{\Delta t} \quad \left(=\frac{\eta+\{ S/h \} }{\Delta t}\right) \qquad \to \qquad \frac{\Delta t}{\Delta T}=\frac{S/h}{\eta_{1}} \label{dilat}
\end{equation}
Note that, because $S/h \ge \eta_{1}$, this last relationship necessarily \emph{implies} that \emph{a clock that does not measure increments of proper time, measures time intervals $\Delta t$ that are dilated compared to the proper ones $\Delta T$}. Let an observer be at rest with a quantum of action. The observer 'eternally' sees '1' with perfect certainty. What is being 'measured' by this observer is physically the same as what another observer measures in a different frame moving with relative velocity $v$ and where $2n$ experiments in the unary radix take place yielding $n$ quanta of action. If all $2n$ experiments \emph{take place simultaneously} then the probability that a certain chain of zeroes and ones is compatible with this situation is given by Eq. (\ref{proba}). Remind that for the observer in the rest frame there are no simultaneous experiments at all, but the single outcome '1' with perfect certainty. Thus, to account for both observers, the one at rest and the one moving with velocity $v$, it is clear that Eq. (\ref{proba}) has to be extended to
\begin{equation}
\frac{1}{2^{2n}}{2n \choose n}\left(\frac{v}{c}\right)^{2n}
\end{equation}
Thus, for $v=c$ Eq. (\ref{proba}) is regained and for $v=0$ (hence $n=0$ experiments) it yields the outcome '1' of the observer at rest. The number of simultaneous experiments in the moving frame can freely fluctuate. Each possible chain of $2n$ simultaneous experiments has the same content in action $\eta_{1}$ since any of these possibilities being the case corresponds to the outcome '1' of the observer at rest. We, thus, obtain
\begin{equation}
\frac{\Delta t}{\Delta T}=\frac{S/h}{\eta_{1}}=\frac{1}{\eta_{1}}\sum_{n=0}^{\infty}p(n)\eta_{1}=\sum_{n=0}^{\infty} \frac{1}{2^{2n}}{2n \choose n}\left(\frac{v}{c}\right)^{2n}=\frac{1}{\sqrt{1-\left(\frac{v}{c}\right)^{2}}} \label{Lorentz}
\end{equation}
where we have used that 
\begin{equation}
\frac{1}{\sqrt{1-4u}}=\sum_{n=0}^{\infty} {2n \choose n}u^{n} \label{cenbin}
\end{equation}
is the generating function of the central binomial coefficients \cite{Koshi} (Eq. (\ref{Lorentz}) then simply follows by taking $u=\left(\frac{v}{2c}\right)^{2}$ in Eq. (\ref{cenbin})). Eq. (\ref{Lorentz}) is the celebrated Lorentz time dilation. Here we have derived it from quantum mechanical probabilistic arguments suggested by the postulate on the physical interpretation of the unary radix, instead of the usual geometric-kinematic ones. 

We can now ask how many experiments are simultaneous on average. We note that, indeed, the sum in Eq. (\ref{Lorentz}) can be interpreted as a grand canonical partition function of the ``spacetime statistics'' presented in Section \ref{statistics} where $v/c$ plays the role of a ``fugacity''. For $v<c$ the average value $<2n>$ of the length of the chains converges to
\begin{equation}
<2n>=\frac{\sum_{n=0}^{\infty} \frac{2n}{2^{2n}}{2n \choose n}\left(\frac{v}{c}\right)^{2n} }{\sum_{n=0}^{\infty} \frac{1}{2^{2n}}{2n \choose n}\left(\frac{v}{c}\right)^{2n}}=\frac{v^{2}}{c^{2}-v^{2}}=\frac{m^{2}}{m_{0}^{2}} \frac{v^{2}}{c^{2}}=\frac{E^{2}v^{2}}{m_{0}^{2}c^{6}}  \label{nav}
\end{equation} 
%=\frac{(E+m_{0}c^{2})^{2}v^{2}}{2m_{0}^{2}c^{6}}
where $m$ is the relativistic mass of the particle, defined as
\begin{equation}
m \equiv \ \frac{m_{0}}{\sqrt{1-\left(\frac{v}{c}\right)^{2}}}=\frac{c}{v}m_{0}\sqrt{<2n>}=\frac{E}{c^{2}}
\end{equation}
The energy is thus consistent with Einstein's relationship
 \begin{eqnarray}
 E &=& mc^{2}=\frac{m_{0}c^{2}}{\sqrt{1-\left(\frac{v}{c}\right)^{2}}} 
 =c\sqrt{p^{2}+m_{0}^{2}c^{2}} \label{enom}
 \end{eqnarray}
where $p\equiv mv$. Eq. (\ref{nav}) diverges when $v=c$. Furthermore, $v$ cannot exceed $c$ since then Eq. (\ref{Lorentz}) would diverge and Eq. (\ref{nav}) would yield negative values. Eq. (\ref{nav}) has a further important consequence: \emph{it directly connects the chains of zeroes and ones of Section \ref{statistics} to physical quantities: The average length of the chains when described in radix $\eta_{1}$ is a function of the ratio $v/c$}.

\begin{figure}
\includegraphics[width=0.4 \textwidth, angle=270]{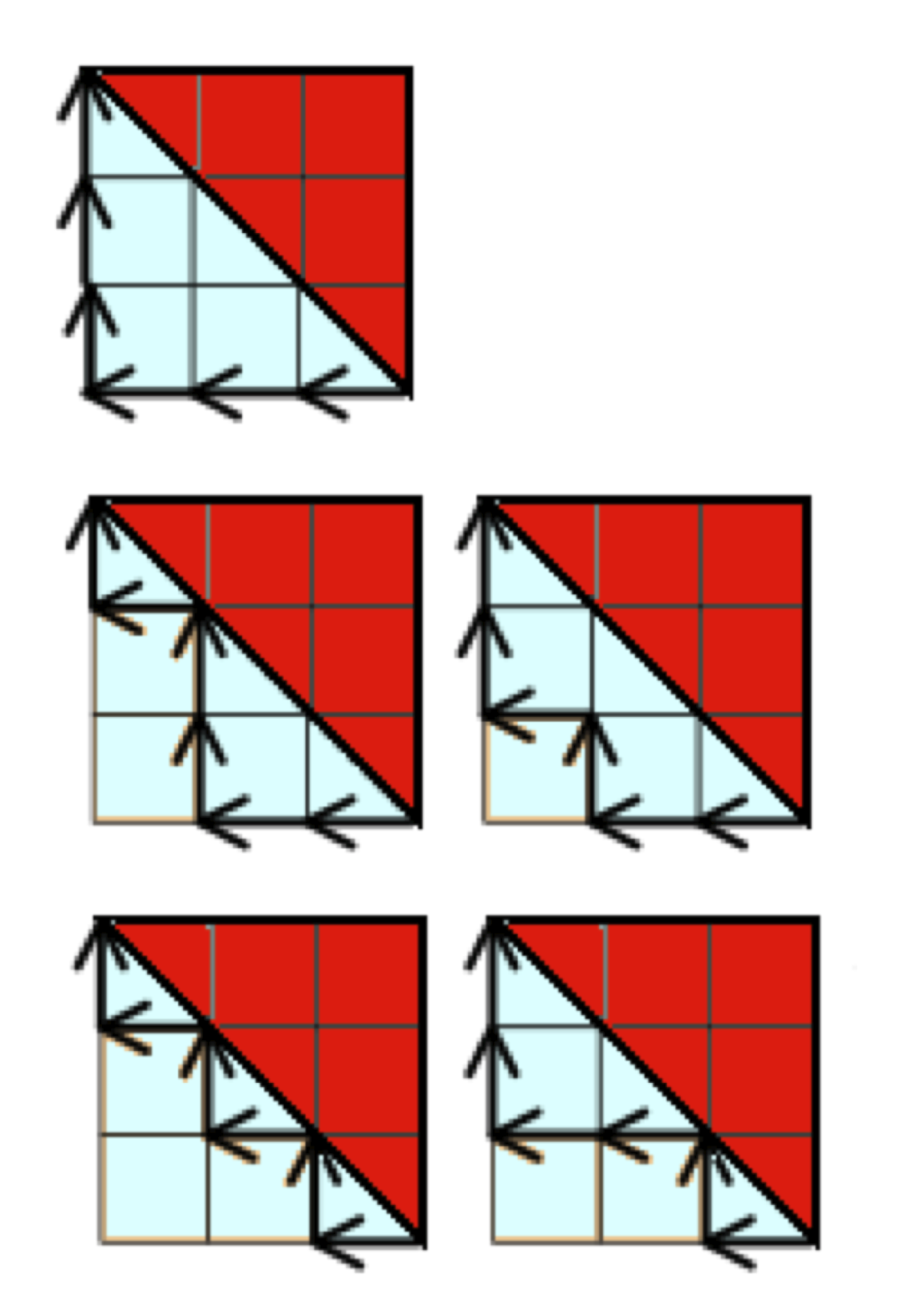}
\begin{center}
\caption{All $C_{3}=5$ possible physical paths in spacetime for the case $\eta=3$ (see text). Space $|\Delta \mathbf{r}|$ is plotted in the abscissa and $c\Delta t$ in the ordinate. Shown is the relevant part of the Minkowski spacetime. The cyan region constitutes the accessible points in spacetime and the red one the inaccessible ones.} \label{paths}
\end{center}
\end{figure}

\emph{Special relativity and relativistic quantum mechanics are consequences of the principle of least radix economy and the physical implications of the unary radix.} We can now finish the picture on familiar grounds. From Eq. (\ref{Lorentz}) we have
\begin{equation}
\left(\frac{\Delta t}{\Delta T}\right)^{2}-\left(\frac{v\Delta t}{c\Delta T}\right)^{2}=1
\end{equation}
from which we obtain, by using that $|\Delta \mathbf{r}|=v \Delta t$
\begin{equation}
c^{2}\left(\Delta t \right)^{2}-|\Delta \mathbf{r}|^{2}=c^{2}\left(\Delta T\right)^{2}\equiv \left(\Delta s\right)^{2}
\end{equation}
where we have defined the length element $\Delta s$. Thus, if all increments are infinitesimally small we obtain
\begin{equation}
c^{2}d t^{2}-|d \mathbf{r}|^{2}=d s^{2}
\end{equation}
which is the length element of the Minkowskian geometry. Now we can understand what the Catalan numbers in Eq. (\ref{supercatal}) count: The chains of zeroes and ones correspond to paths in the forward lightcone and the Catalan number $C_{\eta}$ correspond to all possible paths in spacetime within the reach of $2\eta$ experiments. Experiments with outcome '1' correspond to quanta of action along the time-like worldline while experiments with outcome '0' vacuum states corresponding to propagation along space-like coordinates. Note that for the observer at rest with the quantum of action, the latter kind of propagation is not possible since the observer just measures '1' quantum of action only along the time-like worldline, which corresponds to a proper time since his measurement corresponds to a quantum of energy as well. Note that, from Eqs. (\ref{supercatal}) and (\ref{Lorentz})
\begin{equation}
\frac{S}{h}=\sum_{n=0}^{\infty}p(n)=\sum_{n=0}^{\infty}\frac{(n+1)}{2^{2n}}\left(\frac{v}{c}\right)^{2n}C_{n}
\end{equation}
The sum runs over all possible paths within the forward Minkowski cone. In Fig. \ref{paths} these paths are sketched for the case $n=3$. We have $C_{3}=5$ paths. Space ($|\Delta \mathbf{r}|$) is plotted in the abscissa and time $(c\Delta t)$ in the ordinate. The cyan region contains all points in spacetime that are accessible within the forward Minkowski cone. The red region contains all points that are inaccessible. The paths, from top to bottom and left to right correspond thus to the chains '101010', '101100', '110010','110100', '111000'. Since these chains have length $6$, from Eq. (\ref{nav}) this means that this describes a situation that we would find on average if we take $v/c=\sqrt{6/7}$.

In coordinates, we have
\begin{equation}
d s^{2}=c^{2}d t^{2}-d x^{2}-d y^{2}-d z^{2} \label{Minko}
\end{equation}
We can thus define the quadrivector $(x^{0},x^{1},x^{2},x^{3}) \equiv (ct, x, y, z)$ and express Eq. (\ref{Minko}) as
\begin{equation}
d s^{2}=\eta_{\mu\nu}dx^{\mu}dx^{\nu}   \label{Minko2}
\end{equation}
where we have used Einstein's summation convention for repeated indices and
\begin{equation}
\eta_{\mu\nu}=\eta^{\mu\nu}=\left( \begin{array}{rrrr} 1 & 0 & 0 & 0 \\ 0 & -1 & 0 & 0 \\ 0 & 0 & -1 & 0 \\ 
0 & 0 & 0 & -1 \end{array} \right)
\end{equation}
is the metric tensor for the Minkowski metric. Note also that, if we now define the quadrimomentum $(p^{0},p^{1},p^{2},p^{3}) \equiv (E/c, p_{x}, p_{y}, p_{z})$ Eqs. (\ref{HJ2}) can all be written in a compact way as
\begin{equation}
p_{\mu}=-\partial_{\mu}S \label{HJ3}
\end{equation} 
with $p_{\mu}=\eta_{\mu\nu}p^{\nu}$ and where the label $k$ has been dropped. The following scalar defines the rest mass of the particle 
\begin{equation}
p_{\mu}p^{\mu}=-p_{\mu}\eta^{\mu\nu}\partial_{\nu}S=\eta^{\mu\nu}\partial_{\nu}S\partial_{\mu}S=\frac{E^{2}}{c^{2}}-\mathbf{p}^{2} \equiv m_{0}^{2}c^{2}
 \label{HJ4}
\end{equation} 
from which the energy-momentum relationship, Eq. (\ref{enom}) is again obtained
\begin{equation}
E=c\sqrt{\mathbf{p}^{2}+m_{0}^{2}c^{2}} \label{emom}
\end{equation}

A relativistic wave equation which is first order in time and space can now be derived following well-known steps that involve Clifford algebra, but using the wavefunction Eq. (\ref{genform}) and Eqs. (\ref{HJ3}). Let us first introduce the \emph{matrices} $\gamma^{k}$ through the following relationship
\begin{equation}
    \{\gamma^\mu,\gamma^\nu\} \equiv  \gamma^\mu\gamma^\nu+\gamma^\nu\gamma^\mu= 2 \eta^{\mu\nu} \label{anticoms}
\end{equation}
Such expression defines the Clifford algebra $C\ell_{1,3}(\mathbb{R})$ (up to an unimportant factor of 2). Clifford algebras allow us to take the square root of operators involving derivatives in spacetime \cite{Snygg}. We further define 
\begin{equation}
\slashed p \equiv \gamma^{\mu}p_{\mu}=-\gamma^{\mu}\partial_{\mu}S \equiv-\slashed \partial S
\end{equation}
where we have used Feynman' slash notation to introduce also the operator $\slashed \partial \equiv \gamma^{\mu}\partial_{\mu}$. By using now Eq. (\ref{anticoms}) together with Eq. (\ref{Minko}) we have, from Eq. (\ref{HJ4}), remarkably
\begin{equation}
\slashed p \slashed p = (\slashed \partial S)^{2}\mathbf{1}=p_{\mu}p^{\mu}\mathbf{1}= m_{0}^{2}c^{2}\mathbf{1}
\end{equation}
where $\mathbf{1}$ is the identity matrix. Therefore, we also have
\begin{equation}
\slashed \partial S= \pm m_{0}c\mathbf{1}
\end{equation}
Taking the minus sign of this latter equation and using Eq. (\ref{genform}) and applying the $\slashed \partial$ operator to both sides
\begin{eqnarray}
\slashed \partial \psi &=&\sum_{k=-\infty}^{\infty}\widetilde{\psi}(k)\slashed \partial e^{ikS/\hbar}=\frac{i}{\hbar}\sum_{k=-\infty}^{\infty}k\widetilde{\psi}(k)e^{ikS/\hbar}\slashed \partial S  \nonumber \\
&=& -\frac{im_{0}c\mathbf{1}}{\hbar} \sum_{k=-\infty}^{\infty}k\widetilde{\psi}(k)e^{ikS/\hbar} = -\frac{im_{0}c}{\hbar}\mathbf{1} \hat{K}\psi
\label{genformdi}
\end{eqnarray}
where $\hat{K}$ is the index operator that we defined in Eqs. (\ref{index1}) and (\ref{index2})
We thus obtain the relativistic wave equation
\begin{equation}
i\hbar\gamma^{\mu}\partial_{\mu}\psi-m_{0}c\mathbf{1}\hat{K}\psi=0 \label{genDic}
\end{equation}
In order for this equation to have solutions the wavefunction $\psi$ must be a vector. It indeed defines what is called a \emph{spinor field}. When only the mode $k=1$ is present in the wavefunction, $\psi=\widetilde{\psi}(1)e_{1}$ the latter equation reduces to the Dirac equation
\begin{equation}
i\hbar\gamma^{\mu}\partial_{\mu}\psi-m_{0}c\mathbf{1}\psi=0
\end{equation}
with $\psi$ being a four-component vector, a bispinor. In Dirac representation, the four contravariant gamma matrices are
\begin{eqnarray}
&&    \gamma^0 = \begin{pmatrix} \ 1 \ & \ 0 \ & 0 & 0 \\ 0 & 1 & 0 & 0 \\ 0 & 0 & -1 & 0 \\ 0 & 0 & 0 & -1 \end{pmatrix} \qquad \gamma^1 = \begin{pmatrix} 0 & 0 &\ 0 \ &\  1\ \\ 0 & 0 & 1 & 0 \\ 0 & -1 & 0 & 0 \\ -1 & 0 & 0 & 0 \end{pmatrix} \\
&&    \gamma^2 = \begin{pmatrix} 0 & 0 & 0 & -i\ \\ 0 &\ 0\ &\ i \ & 0 \\ 0 & i & 0 & 0 \\ -i & 0 & 0 & 0 \end{pmatrix} \qquad \gamma^3 = \begin{pmatrix} 0 &\ 0\ &\ 1\ & 0 \\ 0 & 0 & 0 & -1 \\ -1 & 0 & 0 & 0 \\ 0 & 1 & 0 & 0 \end{pmatrix} 
\end{eqnarray}
which can be written in terms of the $2\times 2$ Pauli matrices $\sigma_{i}$ as
\begin{equation}
    \gamma^0 = \begin{pmatrix} I & 0 \\ 0 & -I \end{pmatrix},\quad \gamma^i = \begin{pmatrix} 0 & \sigma^i \\ -\sigma^i & 0 \end{pmatrix} \,  
\end{equation}
where $I$ denotes the $2\times 2$ identity matrix.

We must now justify why $k=1$ describes spin-1/2 states, since we have used this in our derivation of the Dirac equation. Since $\eta_{1}$ describes particles, we must work in this radix in order to be able to interpret spin as an internal property of them, if we stick to our postulates. From Eq. (\ref{genformIIb}) we have, if $\eta=\eta_{1}$, 
\begin{eqnarray}
\psi(S)&=&\sum_{k=-\infty}^{\infty}\widetilde{\psi}(k)e^{i2\pi kS/h}= \sum_{k=-\infty}^{\infty}\widetilde{\psi}(k)e^{i4\pi k\left \{\frac{S}{2h}\right \}} \label{gen8} 
\end{eqnarray}
where we have used Eq. (\ref{eqmode1}). Since $\left \{\frac{S}{2h}\right \}$ is \emph{any} number $\in [0,1)$, it is clear then that there exist two different kinds of states with different periodicities, which depend on whether the quantity
\begin{equation}
s \equiv \frac{k}{2}
\end{equation} 
is integer or half-integer, i.e. whether $k$ is even or odd respectively. We interpret $s$ as the \emph{spin state} of the particle. When $s$ is half-integer valued we say that the particle is a \emph{fermion} and when it is an integer we call it \emph{boson}. We thus see that $k=1$ above describes particles with spin $s=1/2$. If we replace $k=1$ in Eq. (\ref{gen8}) we also see the $4\pi$ periodicity which is typical of the motion of the electron \cite{Snygg}. This justifies having taken $k=1$ to obtain the Dirac equation from the more general equation Eq. (\ref{genDic}). 

If $\eta > \eta_{1}$ then $k$ in Eq. (\ref{genform}) does no longer describe spin states. At the end of Section \ref{noncom} we have seen an example where $k$ takes indeed values between $0$ and $\eta-1$, with $\eta>1$, and we have mentioned their relationship with states of angular momentum. Such states are not spin states because they do not describe a system where the unary radix is the optimal one. The problem of how to describe particles with higher spin by means of Eq. (\ref{genDic}) shall be discussed elsewhere. We conjecture that a connection should  exist between this equation and the Bargmann-Wigner equations.

Our definition of spin is in tune with the interpretation of spin made in algebraic quantum field theory. See, for example Eqs. I.3.26 and I.3.27 in \cite{Haag}. There a ``little Hilbert space'' is introduced depending on a parameter $\varphi \in [0, 2\pi)$ which is mathematically equivalent to the orthonormal base Eq. (\ref{ortonor}), as considered in Eq. (\ref{gen8}) for $\eta=\eta_{1}$. Here $\left \{\frac{S}{2h}\right \}$ plays the same role as $\varphi$ there.

\section{Derivation of Boltzmann's principle}
\label{entropy}

The Second Law of Thermodynamics becomes important when the total number of degrees of freedom $N$ in Eq. (\ref{Euler}) is huge. The Second Law finds an elegant interpretation in Statistical Thermodynamics through Boltzmann's principle, which establishes that the equilibrium thermodynamic entropy $\mathcal{S}_{B}$ for an isolated system given by
\begin{equation}
\mathcal{S}_{B}=k_{B}\ln \Omega \label{Bolz}
\end{equation}
attains its maximum at equilibrium ($k_{B}$ is the Boltzmann constant). $\Omega$ in Eq. (\ref{Bolz}) is interpreted as the space of configurations of a finite system. For a classical Hamiltonian system where energy is conserved one has \cite{Pathria} \cite{Gross}
\begin{equation}
\Omega=\int \frac{d^{3N}\mathbf{p} d^{3N}\mathbf{q}}{h^{3N}N!}\delta\left(E-H(\mathbf{q},\mathbf{p})\right)
\label{omeg}
\end{equation}
This number corresponds to the total number of attainable microstates in the constant energy surface. Eq. (\ref{Bolz}) is connected to the Gibbs canonical ensemble through the Laplace transform of $\Omega$. Why Eq. (\ref{Bolz}) or, equivalently, the Gibbs ensemble, describes indeed thermodynamic equilibrium is a mistery. We quote Ruelle \cite{Ruelle}:  \emph{The problem of why the Gibbs ensemble describes the thermal equilibrium (at least for ``large systems'') [...] is deep and incompletely clarified}.   

From the principle of least radix economy we can now give a new interpretation of Boltzmann's entropy and the Second Law. To see how, let us remark that $\Omega$ is just a huge number corresponding to all possible configurations of the conservative Hamiltonian system: There is also a huge variety of \emph{finite} paths with dimensionless action $S/h$ (and radix $\eta=\lfloor S/h \rfloor$) that are contained in the constant energy surface and we can express the number $\Omega$ in terms of the radix corresponding to any of these paths.  

At equilibrium, the concept of typicality was coined to describe the paths with the largest probability measure. The principle of least radix economy, as we show now, provides both the right expression for the equilibrium entropy and an estimate of the length of the typical paths. To see this, first observe that the number to be expressed now in the radix $\eta=\lfloor S/h \rfloor$ is $\mathcal{A}=\Omega$ which is a radix-independent fixed quantity given by Eq. (\ref{omeg}). Thus, we have, from (b) Postulate \textbf{B}
\begin{eqnarray}
D_{\varepsilon}\mathcal{C}\left(\eta, \Omega \right)&=&D_{\varepsilon}\left(\eta \left \lfloor 1+\log_{\eta}\Omega \right \rfloor\right) \approx D_{\varepsilon}\left(\eta \log_{\eta}\Omega\right) \nonumber \\
&=&  D_{\varepsilon}\left(\frac{\eta}{\ln \eta}\right) \ln \Omega=\frac{d}{d\eta}\left(\frac{\eta}{\ln \eta}\right) \ln \Omega \ D_{\varepsilon} \eta \nonumber \\ 
&=&0 \label{prinB}
\end{eqnarray}
where Eq. (\ref{dcap}) has been used together with the fact that $\Omega$ is very large and so is its logarithm compared to unity, i.e.
\begin{equation}
\mathcal{C}\left(\eta, \Omega \right) \approx \eta \log_{\eta}\Omega=\eta \frac{\ln \Omega}{\ln \eta} \label{approx}
\end{equation}
We see from Eq. (\ref{prinB}) that each least radix path obeying Eq. (\ref{prinA}) (i.e. satisfying $D_{\varepsilon} \eta=0$) is a local minimum. There is, however, a \emph{global} minimum as well, which occurs when 
\begin{equation}
\left.\frac{d}{d\eta}\left(\frac{\eta}{\ln \eta}\right)\right|_{\eta_{min}}=0 \label{prinC}
\end{equation}
The function $x/\ln x$ has a minimum at $x=e$. Therefore, since $\eta=\lfloor S/h \rfloor$ is an integer the minimum is at $\eta_{min}=3 \ (\approx e)$ \cite{Hurst} which gives the typical paths. Such paths in the constant energy surface are characteristic of thermodynamic equilibrium and are tiny because of the effect of thermalization in phase space (i.e the principle of the equipartition of energy) at equilibrium where the global minimum is attained. The constant energy surface is homogeneously filled by an ensemble of paths and those which are the typical ones have the most significant contribution to the average.  

From Eqs. (\ref{Bolz}), (\ref{prinB}) and (\ref{prinC}) we observe that
\begin{equation}
\mathcal{C}\left(\eta_{min}, \Omega \right)=\mathcal{C}\left(3, \Omega \right)=\frac{3}{\ln 3}\ln \Omega=\frac{3}{k_{B}\ln 3}\mathcal{S}_{B}
\end{equation}
which shows how Boltzmann entropy naturally arises from the principle of least radix economy. Furthermore, since from Eq. (\ref{prinC}) we have
\begin{equation}
\frac{k_{B} \mathcal{C}\left(3, \Omega \right)}{\mathcal{S}_{B}} = \frac{3}{\ln 3} \le \frac{\eta}{\ln \eta}
\end{equation}
and then
\begin{equation}
\mathcal{S}_{B} \ge \frac{k_{B} \ln \eta}{\eta} \mathcal{C}\left(3, \Omega \right) = \frac{3 \ln \eta}{\eta \ln 3} k_{B} \ln \Omega \label{cuacua}
\end{equation}
This suggests to define the nonequilibrium path-dependent entropy
\begin{equation}
\mathcal{S}(\eta) \equiv \frac{3 \ln \eta}{\eta \ln 3} k_{B} \ln \Omega \label{neq}
\end{equation}
and thus, we have, from Eq. (\ref{cuacua}) 
\begin{equation}
\mathcal{S}_{B} \ge \mathcal{S}(\eta) 
\end{equation}
where the equality only holds for $\eta=\eta_{min}$ (i.e. at equilibrium). In a nonequilibrium situation, a significant proportion of paths is different to the typical ones. In its evolution to equilibrium, the average characteristic lengths of the possible paths on the constant energy surface changes with time. Finally, a situation is reached when typicality is most relevant and this corresponds to a situation where equipartition of the energy has taken place. Such a situation is described by Boltzmann entropy $\mathcal{S}_{B}=\mathcal{S}(\eta_{min})$ and this entropy is a maximum compared to any other path-dependent nonequilibrium entropy. This is the second law of thermodynamics. Explicitly, the nonequilibrium entropy Eq. (\ref{neq}) is
%\begin{widetext}
\begin{equation}
\mathcal{S}(\eta)=\frac{3k_{B}}{\ln 3}\frac{\ln \lfloor \frac{1}{h}\int_{t_1}^{t_2} Ldt \rfloor }{\lfloor \frac{1}{h}\int_{t_1}^{t_2} L dt \rfloor} \ln \left[\int \prod_{k=1}^{3N} \frac{dq_{k}d\dot{q}_{k}} {h^{3N}N!}\frac{\partial L}{\partial {\dot q_k}}\delta\left(E+L-\sum_i {\dot q_i} \frac{\partial L}{\partial {\dot q_i}}\right)\right]   \label{neq2}
\end{equation}
%\end{widetext}
and can thus be calculated, given any Lagrangian, for any specific path. The paths which satisfy
\begin{equation}
\left \lfloor \frac{1}{h}\int_{t_1}^{t_2} Ldt \right \rfloor=3
\end{equation}
are typical at equilibrium. In a nonequilibrium situation, there are flow structures in phase space with well defined characteristic lengths (if one thinks for example in the coexistence between KAM tori and a chaotic sea in the weakly chaotic regime of a nonlinear Hamiltonian system, the KAM tori define islands of regular motion with well defined characteristic dimensions \cite{Berdichevsky}). These characteristic lengths can be described by corresponding paths in phase space and the entropy associated to any of these paths will be lower than the entropy associated to a typical segment (which provides the characteristic length for a ``thermalized'' path within the chaotic sea).

The arrow of time is given by the fact that, considering a conservative system we have
\begin{equation}
E = - {\partial S \over \partial t} \approx  - h{\partial \eta \over \partial t} 
\end{equation}
in the limit $\eta$ large, from the first of the Eqs. (\ref{HJ}). Then the equation
\begin{eqnarray}
\frac{\partial \mathcal{S(\eta)}}{\partial t}=\frac{\partial \mathcal{S}}{\partial \eta}{\partial \eta \over \partial t} &=&-\frac{\partial \mathcal{S}}{\partial \eta}\frac{E}{h}=-\frac{3E\mathcal{S}_{B}}{h\ln{3}}\frac{d}{d \eta} \frac{\ln \eta}{\eta} \label{trend}
%\nonumber \\ &=&\frac{3E\mathcal{S}_{B}}{h\ln{3}}\left(\frac{\eta}{\ln \eta}\right)^{2}\frac{\partial}{\partial \eta} \frac{\eta}{\ln \eta}
\end{eqnarray}
has a fixed point at the global minimum Eq. (\ref{prinC}) of the radix economy (i.e. at the maximum of the entropy) where the r.h.s. is zero. This global fixed point is trivially stable for a bounded system (for which $E=-|E|$). This gives a trend to equilibrium that is consistent with the  Second Law of Thermodynamics. 

%From a linear stability analysis of Eq. (\ref{trend}) around the equilibrium point we obtain that, close to equilibrium
%\begin{equation} 
%\mathcal{S(\eta)} \sim \left(\mathcal{S(\eta)}_{0}-\mathcal{S}_{B}\right)e^{-t/\tau_{R}}+\mathcal{S}_{B}
%\end{equation}
%where $\tau_{R}$, the relaxation time, is given by
%\begin{equation}
%\tau_{R}=\frac{9h\ln{3}}{|E|\mathcal{S}_{B}}
%\end{equation}

As a final remark we note that the nonequilibrium entropy Eq. (\ref{neq}) is inversely proportional to $\eta/\ln {\eta}$. The prime number theorem (first proved by Hadamard and de la Vall\'ee-Poussin in 1896) \cite{Schroeder} states that the number of prime numbers $\pi(n)$ below a certain natural number $n$ is asymptotically equal to
\begin{equation}
\pi(n) \sim \frac{n}{\ln {n}}
\end{equation}
Thus, we observe that the nonequilibrium entropy is inversely proportional to the total number of prime numbers below the radix $\eta=\lfloor \frac{1}{h}\int_{t_1}^{t_2} L dt \rfloor$: the larger the number of prime numbers, the lower the entropy. Remarkably, the condition $n$ large at which the prime number theorem holds, coincides with the asymptotic limit ($\eta$ large) that we have used in Eq. (\ref{prinB}). 

%What this result seems to imply is that non-typical regions in phase space where most of the paths have an optimal radix that is not a prime number seem to have the larger entropy, i.e. that \emph{there seem to be more available microstates by the fact that a path with $\eta$ nonprime can subdivide into subunits given by the divisors of $\eta$}.

\section{The Parker-Rhodes combinatorial hierarchy and the strength of fundamental interactions}
\label{Parker}

In the approach presented in Sections \ref{statistics} and \ref{unary} we have distinguished between \emph{quanta of action} and strings of information formed by \emph{quanta of information} that can each contain either a quantum of action or a vacuum state. The idea of information transfer processes in terms of binary digits to describe fundamental physics pervades the work of Frederick Parker-Rhodes (1914-1987) \cite{Parker}. He made the major discovery of a \emph{combinatorial hierarchy} \cite{Bastin}, \cite{Parker}, \cite{Noyes} that is a most deep and intriguing result of relevance to us here. Starting from the most elementary configurations, the combinatorial hierarchy constitutes a process in which chains of zeroes and ones are generated giving rise to numerical values that, quite surprisingly, capture the relative interaction strengths of some known fundamental physical forces. There exist some detailed expositions of the combinatorial hierarchy, e.g. by Ted Bastin and Clive Kilmister \cite{Bastin} who also contributed significantly to the topic. The later joint effort of H. Pierre Noyes, John Amson and David McGoveran added further insights of interest to physics \cite{Noyes}.

We close this article showing how the Parker-Rhodes combinatorial hierarchy is physically understandable within the conceptual framework that we have developed here. We now consider general chains of unequal numbers of zeroes and ones, which are meant to be of the type discussed in Section \ref{statistics} and which codify interacting particles for which a wavefunction of the form Eq. (\ref{genform}) can be constructed. Particles that interact through the strong force (quarks), interact also electromagnetically (they bear electric charge) and gravitationally (they bear mass). If $\Omega$ denotes the number of chains that describe a certain interaction between particles with mass we find that $\Omega_{strong} < \dots < \Omega_{electromagnetic} < \Omega_{gravitation}$ since each set of chains includes the previous one as a subset. We now construct hierarchically chains of zeros and ones, excluding always the chain where all digits are zero (the vacuum), by using the elements obtained in the previous step. We consider first the element '1' denoting the quantum of action and '0' denoting the vacuum state. With these two elements we can build three chains '10', '01' and '11' (the vacuum chain '00' is excluded). We start with these three chains to hierarchically construct others. Thus, if $\Omega_{m-1}$ denotes the number of chains constructed at iteration step $m-1$, the number of new chains generated at step $m$ is 
\begin{equation}
\Omega_{m}=2^{\Omega_{m-1}}-1 \label{Mersenne}
\end{equation} 
This sequence gives $3, 7, 127, 2^{127}-1 \approx 1.7\cdot 10^{38}$ (Sloane A00584, itself a subsequence of Sloane A007013). Eq. (\ref{Mersenne}) constitutes the iteration of the so-called Mersenne operator, which is used in the investigation of prime numbers. This sequence also coincides with the Catalan sequence (Sloane A180094) which gives the total number of steps to reach 0 or 1 starting with $\Omega_{m}$ and applying the map $\Omega_{m} \to$ (number of 1's in the binary expansion of $\Omega_{m}$) repeatedly. 

%170141183460469231731687303715884105727

The \emph{cumulative sum} of all numbers of chains $\sum_{k=0}^{m}\Omega_{k}$ gives the sequence: $3, 10, 137,  2^{127}+136 \approx 1.7\cdot 10^{38}$. This sequence is the central result of Parker-Rhodes theory. Since it is not yet included in the comprehensive Sloane's Online Encyclopedia of Integer Sequences, http://oeis.org (as of 13th October 2014) it can be happily ``baptized'' as \emph{Parker-Rhodes sequence}.

%170141183460469231731687303715884105864

We have now, of course $\Omega_{0} <\Omega_{0}+\Omega_{1} < \dots < \sum_{k=0}^{m}\Omega_{k}$ mimicking the above relationship described for the different physical interactions. We can now define the \emph{strength} of an interaction quite naturally as the ratio of one quantum of action to the total number of chains cumulatively created at step $m$ as
\begin{equation}
\frac{\eta_{1}}{\sum_{k=0}^{m}\Omega_{k}} \label{strength}
\end{equation} 
We have proceeded in a similar way above when deriving the kinetic energy term of special relativity, since we have attributed to each chain of zeros or ones a path in the forward Minkowski cone, representing the propagation of a free particle. Now, the quantum of action is distributed within all possibilities of the interaction potential. The larger the number of possibilities, the lower the action corresponding to each of them. From probabilistic considerations, the meaning of this equation is also clear if we think in terms of a Boltzmann equal a priori probability postulate (which leads to the Boltzmann entropy discussed in Section \ref{entropy} when the same probability is attributed to every microstate in phase space). The total numbers of chains are total numbers of possibilities (\emph{microstates}) of distributing one quantum of action in the configuration space of chains of zeroes and ones. The larger is such number, the larger also the number of microstates and the lower the probability that the quantum of action goes to a particular microstate. The lower is thus the field strength (which is described by this elementary transfer of information). For the first few iterations of the Parker-Rhodes sequence we obtain from Eq. (\ref{strength}) starting with $\Omega_{0}=3$ (corresponding to the elements '00' and '01' and '10') the values
\begin{equation}
\frac{1}{3},\ \frac{1}{10},\ \frac{1}{137},\ \frac{1}{1.7\cdot 10^{38}} \label{PR sequence}
\end{equation} 
Parker-Rhodes gave arguments of why the sequence must finish at the fourth iteration \cite{Bastin}, \cite{Parker}, \cite{Noyes}: It becomes then impossible to construct enough linear operators that discriminate between the chains constructed. Most remarkably, if the strong force is taken to have strength 1, the third and fourth terms of the expansion give the right orders of magnitude of the relative strength of the electromagnetic and the gravitational forces between two protons, respectively. In fact the values approximate the actual \emph{experimental} ones to an error of \emph{less than one thousandth!}

\section{Conclusions}
\label{disc}

In this article we have presented a variational method: the principle of least radix economy, Eq. (\ref{prin}) that has led us to a new interpretation of both classical and quantum mechanics. The dimensionless integer quantity $\eta = \left \lfloor S/ h \right \rfloor$ has been ascribed to the most efficient radix in which numbers entering in physical laws are expressed. Minimizing the radix economy $\mathcal{C}(\eta, \mathcal{A})$ (where $\mathcal{A}=S/h$ is the physical action or a radix-independent quantity), has been shown to yield two different classes of solutions: least action paths and quantum wavefunctions. The Hilbert space of quantum mechanics, the Schr\"odinger equation, and Heisenberg uncertainty relationships have been derived with this principle. The breaking of the commutativity of spacetime geometry and the existence of quantum numbers has then also been elucidated. A new derivation of Lorentz time dilation and Einstein's special relativity has then been accomplished from statistical arguments (instead of the traditional geometric-kinematic approach). We have also derived relativistic wave equations governing the spatiotemporal evolution of spin fields. The radix economy $\mathcal{C}(\eta, \mathcal{A})$ is an entropy-like quantity and we have also shown how classical statistical thermodynamics is encompassed by the principle of least radix economy. 

The central idea of this article is simple, although unfamiliar, and is detailed in Section \ref{radie}. A physical number is not only just a number accompanied by physical units. The number is also represented in a certain radix (a fact that has previously been overlooked even when physicists always like to speak about ``orders of magnitude'') and we claim that this radix is physically important, even when physical laws \emph{at a certain scale} (i.e. when the dynamical variables do not change in many orders of magnitude) are \emph{not} affected by how actual numbers are indeed represented. In the quest for a unified theory of physics, we propose that the radix change provides the necessary degree of freedom (\emph{not} a free parameter) to bring physics at all scales together. We also propose that nature dynamically exploits this radix change and that this explains the wave-particle duality found in the quantum realm (and the least action principle in the classical realm). We claim that the quantum of action (naturally described by the unary radix) is the main building block in defining any physical action (as we have shown through the explicit construction of the kinetic energy term). We thus suggest that there exists a ``radix relativity'' that concerns the physical action and which, together with the conservation laws related to the symmetries of the action itself through Noether's theorem, needs to be accounted for in order to better understand the wide variety of dynamical behavior found in the universe (and why classical determinism and quantum mechanics \emph{coexist} within the same physical reality).

\begin{acknowledgements}
I thank Prof. Jos\'e Antonio Manzanares for his many helpful remarks and Prof. Jos\'e Mar\'ia Isidro San Juan for his  comments on a previous version of this manuscript. Past support from the Technische Universit\"at M\"unchen - Institute for Advanced Study (funded by the German Excellence Initiative) through a three-years Carl von Linde Junior Fellowship (when this research was initiated) is also gratefully acknowledged.
\end{acknowledgements}


\begin{thebibliography}{}

\bibitem{Wolfram} Wolfram, S.:  A New Kind of Science. Wolfram Media Inc., Champaign IL, (2002)
\bibitem{Fredkin} 
Fredkin, E.: Digital mechanics. Physica D  45,  254-270 (1990)
\bibitem{McCauley} McCauley, J. L.: Chaos, Dynamics and Fractals: an algorithmic approach to deterministic chaos. Cambridge University Press, Cambridge UK, (1993)
\bibitem{Turing}
A. M. Turing: Systems of Logic Based on Ordinals. Ph.D. thesis, Princeton University, (1939), p. 8.
\bibitem{Hooft}
t'Hooft, G.: Quantum Gravity as a Dissipative Deterministic System. Class. Quant. Grav. 16 3263-3279 (1999)
\bibitem{Bastin}
Bastin, T., Kilmister, C. W.: Combinatorial Physics. World Scientific, Singapore (1995)
\bibitem{Parker}
Parker-Rhodes, A. F.: The theory of indistinguishables. D. Reidel, Dordrecht (Holland) (1981)
\bibitem{Noyes}
Noyes, H. P.: Bit-String Physics. World Scientific, Singapore (2001)
\bibitem{Kauffman1}
Kauffman, L. H.: Non-commutative worlds. New J. Phys. 6, 173 (2004)
\bibitem{Kauffman2}
Kauffman, L. H., Noyes, H. P.: Discrete Physics and the Dirac Equation. Phys. Lett. A, 218, 139-146 (1996)
\bibitem{MC2}
McCauley, J. L.: Chaotic Dynamical Systems as Automata, Z. Naturforsch. A  42  547-555 (1987)
\bibitem{MCPAL1}
McCauley, J. L.,  Palmore, J. I.: Computable chaotic orbits. Phys. Lett. A 115, 433-436 (1986)
\bibitem{comphys}
Garcia-Morales, V.: Nonlocal and global dynamics of cellular automata: A theoretical computer arithmetic for real maps, arXiv:1312.6534 [math-ph] (2013)
\bibitem{VGM1} Garcia-Morales, V.: Universal map for cellular automata. Phys. Lett. A 376, 2645-2657 (2012)
\bibitem{VGM2} Garcia-Morales, V.: Symmetry analysis of cellular automata. Phys. Lett. A 377  276-285 (2013)
\bibitem{VGM3} Garcia-Morales, V.: Origin of complexity and conditional predictability in cellular automata. Phys. Rev. E 88, 042814 (2013)
\bibitem{VGM4} Garcia-Morales, V.: Universal map for substitution systems, arXiv:1309.5254 [math-ph] (2013) 
\bibitem{Wigner}
Wigner, E. P.: The unreasonable effectiveness of mathematics in the natural sciences. Comm. Pure Appl. Math. 13, 1-14 (1960)
\bibitem{Knuth} 
Knuth, D. E.: The Art of Computer Programming vol. II: Seminumerical Algorithms (3rd edition). Addison Wesley, Reading MA (1998), p. 319
\bibitem{Hurst} 
Hurst, S.L.: Multiple-Valued Logic: its Status and its Future. IEEE Trans. Computers, C33, 1160-1179 (1984) 
\bibitem{Schroeder}
Schroeder, M.: Number Theory in Science and Communication. Springer Verlag, Berlin (2009).
\bibitem{Dirac}
Dirac, P. A. M.: The Principles of Quantum Mechanics. Oxford University Press, Oxford, UK (1988) 
\bibitem{Neumann}
von Neumann, J.: Mathematical Foundations of Quantum Mechanics. Princeton University Press, Princeton, (1996). 
\bibitem{Tits} 
Titchmarsh, E. C., Heath-Brown, D. R.: The Theory of the Riemann Zeta-function. Oxford University Press, Oxford, UK (1986), p. 15, Eq. 2.1.7
\bibitem{Dym}
Dym, H., McKean, H. P.: Fourier Series and Integrals. Academic Press, London (1972).
\bibitem{Lanczos}
Lanczos, C.: Linear Differential Operators. Van Nostrand, London (1961).
\bibitem{Heisenberg}
Heisenberg, W.: The Physical Principles of the Quantum Theory. Dover, New York (1949). 
\bibitem{Niven}
Niven, I.: Irrational numbers. The Mathematical Association of America (Carus Mathematical Monographs No. 11), Washington (1956). 
\bibitem{Bohm}
Bohm, D.: Quantum Theory. Dover, New York (1979). 
\bibitem{Tanner}
Tanner, G., Richter, K., Rost, J.M.: The theory of two-electron atoms: Between ground state and complete fragmentation. Rev. Mod. Phys. 72, 497 (2000). 
\bibitem{Gutzwiller}
Gutzwiller M. C.: Chaos in Classical and Quantum Mechanics, Springer Verlag, New York (1990). 
\bibitem{Zygmund}
Zygmund, A..: Trigonometric series. Cambridge University Press, Cambridge (1959).
\bibitem{Weyl}
Weyl, H.: The Theory of Groups and Quantum Mechanics. Dover, New York (1950).
\bibitem{Akhiezer}
Akhiezer, N. I. and Glazman, I. M.: Theory of linear operators in Hilbert space. Dover, New York (1961).
\bibitem{Streater}
Streater, R. F., Wightman, A. S.: PCT, Spin and Statistics, and all That. W. A. Benjamin, New York (1964).
\bibitem{Halmos}
Halmos, P. R.: Introduction to Hilbert Space and the Theory of Spectral Multiplicity. Chelsea, New York (1951).
\bibitem{HooftNEW}
t'Hooft, G.: The Cellular Automaton Interpretation of Quantum Mechanics. arXiv:1405.1548v2 (2014)
\bibitem{Bruckner}
Bruckner, B. S., Bruckner, J. B., Thomson, A. M.: Real Analysis. Prentice Hall, Upper Saddle River NJ (1996).
\bibitem{Tinkham}
Tinkham, M.: Group Theory and Quantum Mechanics. Dover, New York (1964).
\bibitem{Feynman} 
Feynman, R. P..: Space-Time Approach to Non-Relativistic Quantum Mechanics. Rev. Mod. Phys. 20, 367-387.
\bibitem{Kennard}
Kennard, E. H.: Zur Quantenmechanik einfacher Bewegungstypen. Z. Phys. 44, 326-352 (1927).
\bibitem{Robertson}
Robertson, H.P.: The uncertainty principle. Phys. Rev. 34, 163-164 (1929)
\bibitem{Connes}
Connes, A.: Noncommutative Geometry. Academic Press (1994).
\bibitem{Chevalley}
Chevalley, C.: Sur certains groupes simples. Tohoku Math. J. (2), 7, 14-66 (1955)
\bibitem{Carter}
Carter, R. W.: Simple groups of Lie type. John Wiley \& Sons (1989).
\bibitem{Schwinger}
Schwinger, J.: Unitary operator bases. Proc. Natl. Acad. Sci. USA 46, 570-579 (1960)
\bibitem{Kac}
Kac, M.: Statistical independence in probability, analysis and number theory. The Mathematical Association of America (Carus Mathematical Monographs No. 12), Washington (1959).
\bibitem{VGMStat}
Garcia-Morales, V; Pellicer, J.: Microcanonical foundation of nonextensivity and generalized thermostatistics based on the fractality of the phase space. Physica A, 361, 161-172 (2006). 
\bibitem{Annals}
Garcia-Morales, V.; Pellicer, J.; Manzanares, J. A.: Thermodynamics based on the principle of least abbreviated action: Entropy production in a network of coupled oscillators. Ann. Phys. (New York), 323: 1844-1858 (2008). 
\bibitem{deBroglie1}
de Broglie, L. La Thermodynamique de la particule isol\'e. Gauthier-Villars, Paris (1964). 
\bibitem{deBroglie2}
de Broglie, L. The reinterpretation of wave mechanics. Found. Phys. 1, 5-15 (1970). 
\bibitem{Koshi}
Koshi, T.: Triangular arrays with applications. Oxford University Press (2011).
\bibitem{Snygg}
Snygg, J.: A New Approach to Differential Geometry Using Clifford's Geometric Algebra. Springer, New York (2012).
\bibitem{Haag}
Haag, R. Local Quantum Physics: Fields, Particles and Algebras. Springer, Berlin (1996). 
\bibitem{Pathria}
Pathria, R. K., Beale, P. D.: Statistical Mechanics. Elsevier, Amsterdam (2011) 
\bibitem{Gross}
Gross, D. H. E.: Microcanonical thermodynamics. World Scientific, Singapore (2001) 
\bibitem{Ruelle} 
Ruelle, D.: Thermodynamic formalism. Cambridge University Press, Cambridge UK, 2nd Edition, (2004)  p. 4.
\bibitem{Berdichevsky} 
Berdichevsky, V. L.: Thermodynamics of chaos and order. Addison Wesley Longman, Essex, UK (1997).
\end{thebibliography}
\end{document}